\documentclass[lettersize,journal]{IEEEtran}
\usepackage{amsmath,amsfonts}
\usepackage{algorithmic}
\usepackage{algorithm}
\usepackage{array}
\usepackage[caption=false,font=normalsize,labelfont=scriptsize,textfont=sf]{subfig}
\usepackage{textcomp}
\usepackage{stfloats}
\usepackage{url}
\usepackage{verbatim}
\usepackage{graphicx}
\usepackage{cite}
\usepackage{bm}
\usepackage{color}
\usepackage{subfloat}
\usepackage{makecell}
\usepackage{amssymb}
\usepackage{booktabs}
\usepackage{microtype}
\begin{document}
\title{Dynamic Precoding for Near-Field\\Secure Communications: Implementation\\and Performance Analysis}
\author{Zihao Teng, \IEEEmembership{Graduate Student Member, IEEE}, Jiancheng An, \IEEEmembership{Member, IEEE}, Christos Masouros, \IEEEmembership{Fellow, IEEE}, Hongbin Li, \IEEEmembership{Fellow, IEEE}, Lu Gan, \IEEEmembership{Member, IEEE}, and Derrick Wing Kwan Ng, \IEEEmembership{Fellow, IEEE}
\thanks{This work is supported by National Natural Science Foundation of China 62471096. The work of H. Li is supported in part by the National Science Foundation under grants CCF-2316865, ECCS-2212940, and ECCS-2332534. This paper has been presented in part at the IEEE Global Communications Conference (GLOBECOM), Cape Town, South Africa, 2024 \cite{glb}. \emph{(Corresponding author: Jiancheng An.)}}
\thanks{Z. Teng and L. Gan are with the School of Information and Communication Engineering, University of Electronic Science and Technology of China (UESTC), Chengdu, 611731, China. L. Gan is also with the Yibin Institute of UESTC, Yibin, Sichuan, 643000, China (e-mail: tzhuestc@163.com; ganlu@uestc.edu.cn).}
\thanks{J. An is with the School of Electrical and Electronics Engineering, Nanyang Technological University (NTU), Singapore 639798 (e-mail: jiancheng.an@ntu.edu.sg).}
\thanks{C. Masouros is with the Department of Electronic and Electrical Engineering, University College London, WC1E 7JE London, U.K. (e-mail: c.masouros@ucl.ac.uk).}
\thanks{H. Li is with the Department of Electrical and Computer Engineering, Stevens Institute of Technology, Hoboken, NJ 07030, USA (e-mail: hli@stevens.edu).}
\thanks{D. W. K. Ng is with the School of Electrical Engineering and Telecommunications, University of New South Wales (UNSW), Sydney, NSW 2052, Australia (e-mail: w.k.ng@unsw.edu.au).}}
\markboth{DRAFT}{DRAFT}
\maketitle
\begin{abstract}
The increase in antenna apertures and transmission frequencies in next-generation wireless networks is catalyzing advancements in near-field communications (NFC). In this paper, we investigate secure transmission in near-field multi-user multiple-input single-output (MU-MISO) scenarios. Specifically, with the advent of extremely large-scale antenna arrays (ELAA) applied in the NFC regime, the spatial degrees of freedom in the channel matrix are significantly enhanced. This creates an expanded null space that can be exploited for designing secure communication schemes. Motivated by this observation, we propose a near-field dynamic hybrid beamforming architecture incorporating artificial noise, which effectively disrupts eavesdroppers at any undesired positions, even in the absence of their channel state information (CSI). Furthermore, we comprehensively analyze the dynamic precoder's performance in terms of the average signal-to-interference-plus-noise ratio, achievable rate, secrecy capacity, secrecy outage probability, and the size of the secrecy zone. In contrast to far-field secure transmission techniques that only enhance security in the angular dimension, the proposed algorithm exploits the unique properties of spherical wave characteristics in NFC to achieve secure transmission in both the angular and distance dimensions. Remarkably, the proposed algorithm is applicable to arbitrary modulation types and array configurations. Numerical results demonstrate that the proposed method achieves approximately 20\% higher rate capacity compared to zero-forcing and the weighted minimum mean squared error precoders.
\end{abstract}


\section{Introduction}
\IEEEPARstart{T}{he} sixth-generation (6G) networks are anticipated to support heterogeneous applications, including extended reality (XR), comprehensive Internet-of-Things (IoT) services, and multi-sensory communications \cite{huawei,an2024emergingtechnologiesintelligentmetasurfaces}. Consequently, they will rely on innovative wireless technologies to provide ultra-fast, reliable, and massive connectivity \cite{2023arXiv230909242A,anfim2025}. To significantly enhance the channel capacity in multiple-input multiple-output (MIMO) systems, the deployment of extremely large-scale antenna arrays (ELAA) is crucial for advancing next-generation wireless networks \cite{2023arXiv230909242A,9903389,10220205,10130641,wang2024tuto, VTC_2024_Teng_Low}. Moreover, terahertz frequency bands offer abundant unlicensed bandwidth resources, facilitating high data rates and transmissions with minimal latency. In particular, by applying ELAA in terahertz frequency bands, the Fraunhofer distance, which differentiates near- and far-field regions, will increase to the order of hundreds of meters \cite{2023arXiv230909242A}. For instance, a square array measuring $0.5$ m $\times$ $0.5$ m operating at a carrier frequency of $60$ GHz has a Fraunhofer distance of up to 200 meters. As a result, in next-generation wireless communications, signal transmission predominantly occurs within the near field \cite{lu2024tuto,wang2024tuto}, rendering the traditional plane wave propagation model invalid. Instead, the spherical wave model becomes essential for accurately depicting wave propagation in these conditions.

In practice, spherical wavefronts facilitate beam focusing in the near field \cite{9738442, MeditCom_2024_Nor_Position, VTC_2024_Jia_Stacked}, enabling beams to be precisely targeted both in the angular and distance dimensions, in contrast to the singular directional steering in far-field scenarios. By leveraging this unique property, multi-user communications within the same angular direction can be realized. Inspired by these advantages, several advanced techniques have been recently developed to achieve near-field MIMO communications \cite{9738442,10365224,10103817,10515204, 10123941,9952197,10158690,10566874}. For example, the authors of \cite{9738442} considered the problem of maximizing the sum rate of multiple users under different antenna architectures, such as fully-digital antenna systems, phase shifter-based hybrid architectures, and dynamic metasurface antenna architecture. Lu \emph{et al.} introduced a hierarchical beam training method for near-field ELAA \cite{10365224}. Wei \emph{et al.} \cite{10103817} investigated the utilization of triple polarization for multi-user wireless communication systems relying on holographic MIMO surfaces and operating in the near-field regime. Furthermore, the authors of \cite{10123941} proposed the concept of location division multiple access in the near field to enhance spectral efficiency beyond conventional spatial division multiple access. Moreover, the authors of \cite{xie2023training, 9913211} studied beam training methods in the near field, while \cite{9957130, 9536436} addressed the beam-splitting effect in wideband NFC systems. Additionally, Zhi \emph{et al.} \cite{zhi2024per} presented a framework for analyzing and designing low-complexity extremely large-scale MIMO systems in the near field with spatial non-stationarities.

 \begin{table*}[]
	\centering
 \renewcommand{\arraystretch}{1.5}
	\caption{Comparison of this work with existing PLS techniques}
	\label{table1}
	\begin{tabular}{|l||l|l|c|c|c|}
		\hline
		PLS techniques& Security dimension & \# of Users & {\makecell[c]{Eavesdroppers' \\CSI requirement}} &{\makecell[c]{Computational \\complexity}} &{\makecell[c]{Security performance \\analysis}} \\ \hline
		Far-field beamforming \cite{6781609,7070667,9913501,9737364} &Angular & Single/Multiple & Yes & High& No \\ \hline
		Far-field DM \cite{9233409,8103767} &Angular & Single/Multiple & No & Low &No \\ \hline
		Near-field beamforming \cite{10436390,10504668} &Angular and distance & Single/Multiple & Yes & High & No \\ \hline
		Near-field DM \cite{10480457} &Angular and distance & Single
		& No & Low & No\\ \hline 
		Proposed scheme &Angular and distance &Multiple&No& Low & Yes \\ \hline
	\end{tabular}
\end{table*}
Additionally, native trustworthiness is an important pillar in 6G communications \cite{9755276}, which requires effective physical layer security (PLS) techniques to ensure secrecy in communication systems \cite{10227884,6817545, TIFS_2025_Niu_On}. In general, supposing the eavesdroppers' channel state information (CSI) is known, the base station (BS) is capable of designing appropriate beamformers to maximize the secrecy rate of the legitimate users or to minimize the received signal power of eavesdroppers \cite{6781609,7070667,9913501,9737364}. However, eavesdroppers are generally passive and non-cooperative in practice, posing a significant challenge in obtaining their accurate CSI at the BS. As a remedy, directional modulation (DM) schemes were proposed to ensure that the desired constellation only appears in the intended direction, while distorting the constellations elsewhere \cite{8839973,9233409,9269392,9790835}. For instance, the concept of artificial noise (AN) \cite{8103767} was introduced to facilitate DM. Specifically, AN is injected into the orthogonal subspace of the vectors spanned by the channels of the legitimate users, thereby distorting only the signals received by eavesdroppers. Despite its efficacy, the AN technique requires extra radio frequency (RF) chains and power consumption. To overcome these limitations, \cite{6544472} proposed an antenna subset modulation (ASM) method for secure millimeter wave communications. Specifically, ASM can randomly select a portion of antennas to transmit signals with modulated beampattern, randomizing the constellations in unintended directions. Building upon this, a new transmit antenna architecture, named programmable weight phased array (PWPA) \cite{8329405}, was proposed, featuring flexible modulation types and rapid switching capabilities.

To further enhance security in wireless communications, dynamic symbol-level precoding (SLP) has been proposed to fully utilize both the CSI and data symbols of the intended users \cite{TWC_2023_An_Fundamental, 9035662}. Specifically, \cite{9053269,9068286} developed SLPs that not only exploit the multi-user interference to guarantee the quality of service for all legitimate transmissions, but also ensure security against eavesdropping. Furthermore, \cite{9149434} presented a multi-beam SLP scheme based on frequency diverse array (FDA), while \cite{9865228} investigated a reconfigurable intelligent surface-aided SLP design for secure wireless transmission. However, these solutions suffer from high computational complexity. In response, more recent studies have introduced low-complexity dynamic precoding methods for achieving PLS. For instance, the authors of \cite{8855017} proposed an analog time-varying precoder design algorithm for a single legitimate user based on polygon construction. This method involves solving the constant modulus phase equation by utilizing a low-complexity polygon construction, with the time-varying solution derived by randomly selecting a polygon from the feasible inner angle solutions. Furthermore, \cite{9690054} presented a multi-user multiple-input single-output (MU-MISO) DM algorithm for a dual-phase shifter array by leveraging the Kronecker decomposition of the linear phase far-field channel.

Despite various efforts, existing SLP techniques have primarily been developed for far-field scenarios. As such, communication security cannot be guaranteed when legitimate users and eavesdroppers are located in the same direction relative to the BS. Meanwhile, due to the broadcasting nature of wireless channels, near-field communication (NFC) also encounters significant security risks. In particular, the open-air transmission inherent in wireless communication makes NFC systems vulnerable to security threats. Fortunately, thanks to the unique properties of near-field channels, secure transmission can be realized by focusing the desired signals on the exact position of the legitimate users, which cannot be achieved by far-field PLS techniques. Inspired by this, most recently, \cite{10436390,10504668} employed the hybrid and analog beamforming architecture to realize secure NFC. However, these approaches still rely on the knowledge of the eavesdroppers' CSI at the BS, leaving it vulnerable to passive eavesdroppers. Moreover, \cite{10480457} developed a near-field DM system against passive eavesdroppers, but it only supports a single legitimate user.

Against this background, this paper presents an AN-based dynamic hybrid precoder design algorithm for secure multi-user NFC. Compared to our conference paper \cite{glb}, i) we propose a hybrid beamforming architecture that simplifies transmitter hardware design. ii) We extend the proposed approach to multi-path channels. iii) We derive the probability density functions of the received signal-to-interference-plus-noise ratio (SINR) for both users and eavesdroppers, and further provide the secrecy outage probability and secrecy zone. iv) We investigate the impact of numbers of paths, antennas, and RF chains on secrecy performance, and present secrecy maps in the simulations. More specifically, the main contributions of this paper are summarized as follows:
\begin{enumerate}
\item In contrast to the prior works which applied AN in far-field channels, we leverage the spherical wavefronts offered by the near-field channel. Thus, secure transmission can be realized between the BS and the exact position of the legitimate users. Furthermore, ELAA, equipped with a significantly larger number of antennas compared to the number of users, provides an expanded null subspace, making the AN-based method particularly suitable for near-field multi-user scenarios. This enables secure transmission with arbitrary modulation types.
\item The dynamic precoders developed are capable of maintaining a constant beamforming gain, effectively eliminating inter-user interference for legitimate users, while generating distorted constellations at other positions in the near-field region. This approach obviates the need to probe the actual CSI of the eavesdroppers. Furthermore, the proposed method possesses low computational complexity and can be applied to the propagation scenario with multiple scatterers in the near field. 
\item We determine the statistical parameters of the AN to satisfy the power constraint and derive the distribution to generate the random initial precoders at different time slots. Based on this distribution, a thorough theoretical analysis of the proposed method is presented, including the average SINR, achievable rate, secrecy capacity, secrecy outage probability, and the size of the secrecy zone.
\end{enumerate}

Extensive simulation results have validated the performance and theoretical derivations of the proposed algorithm, demonstrating its effectiveness and reliability in secure wireless communications within near-field scenarios. Unlike far-field secure transmission schemes that only ensure security in the angular dimension, the proposed algorithm achieves secure transmission in both the angular and distance dimensions. Furthermore, compared to traditional zero-forcing (ZF) and weighted minimum mean squared error (WMMSE) precoders, the proposed method achieves approximately 20\% higher secrecy rates. Additionally, in comparison with existing PLS techniques, the main contributions of this paper are summarized in Table \ref{table1}.

\emph {Notations}: The bold upper- and lower-case letters are adopted to represent matrices and vectors.
In particular, $\left(\cdot\right)^\top$, ${\left(\cdot\right)^*}$, ${\left(\cdot\right)^H}$, ${\left(\cdot\right)^{-1}}$, and ${\left(\cdot\right)^\dagger}$ denote the transpose, conjugate, conjugate transpose, inverse, and Moore-Penrose inverse operators, respectively. $\text{tr}{\left( \cdot \right)}$ is the trace of a matrix, $\text{diag}{\left( \cdot \right)}$ represents a square matrix with the vector's elements on the main diagonal. $\Re\left( \cdot \right)$ and $\Im\left( \cdot \right)$ denote the real part and imaginary part of a matrix, respectively, $\angle\left( \cdot \right)$ is the argument of a complex number. The absolute value, $l_2$ norm, and Frobenius norm are denoted by $\left| \cdot \right|$, $\left\| \cdot \right\|$, and $\left\| \cdot \right\|_F$, respectively. The sets of ${N \times {M} }$ real matrices and $N$-dimensional real vectors are denoted by ${\mathbb{R}^{N \times {M} }}$ and ${\mathbb{R}^{N \times {1} }}$, respectively, while the sets of ${N \times {M} }$ complex matrices and $N$-dimensional complex vectors are denoted by ${\mathbb{C}^{N \times {M} }}$ and ${\mathbb{C}^{N \times {1} }}$, respectively. The statistical expectation is written as ${\mathbb{E}\left[ \cdot \right]}$. $\mathcal{CN}(\bm{\mu},\mathbf{\Sigma})$ and $\mathcal{N}(\bm{\mu},\mathbf{\Sigma})$ denote the complex and real Gaussian distribution with mean $\bm{\mu}$ and covariance matrix $\mathbf{\Sigma}$, respectively. $\mathcal{U}(a,b)$ is adopted to denote the uniform distribution in the interval $(a,b)$.

\section{Near-Field Channel Model and Problem Formulation}
\begin{figure}[!t]
	\centering
	\includegraphics[width=0.5\textwidth]{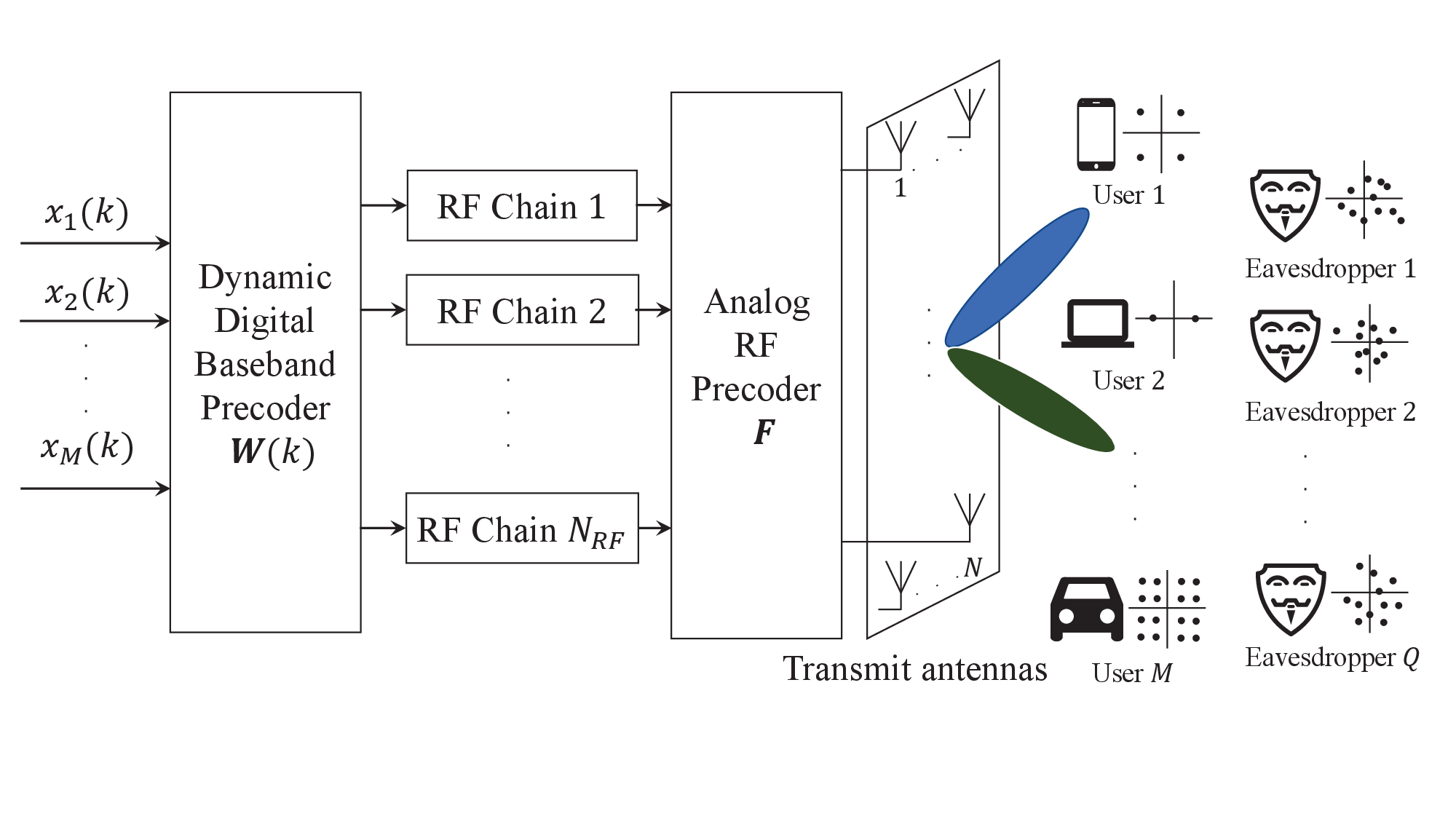}
	\caption{The illustration of a downlink MU-MISO system with multiple eavesdroppers, where $k$ represents the $k$th time slot.}
	\label{illustration}
\end{figure}

We consider a downlink MU-MISO system, as illustrated in Fig. \ref{illustration}, where dynamic digital precoding is performed at the baseband, while analog precoding is applied in the RF domain, forming a hybrid beamforming architecture. The BS is equipped with a planar antenna array. The total number of antenna elements is $N = N_d \times N_e$, where $N_e$ and $N_d$ denote the number of elements in the horizontal direction and the vertical direction, respectively. The $l$th element of the $i$th row is located in three-dimensional Cartesian coordinate system at ${\bf{s}}_{i,l} = [x_l, y_i, 0]^\top$, $ \forall i \in \{1,2,...,N_d\}$, $ \forall l\in \{1,2,...,N_e\}$. The number of RF chains is $N_\text{RF}$. Assume that there are $M$ single-antenna legitimate users located at coordinates ${\bf{r}}^u_m = [x^u_m, y^u_m, z^u_m]^\top$, $\forall m\in \{1,2,...,M\}$. Different from the general uniform spherical wave (USW) model \cite{277864}, we adopt the non-uniform spherical wave (NUSW) model which characterizes the near-field channel more precisely \cite{10220205}. The channel spanning from the BS to the $m$th legitimate user can be expressed as
\begin{align}
\begin{split}
	{{\bf{h}}}({\bf{r}}^u_m) &= \left[ {{A_{1,1}}({{\bf{r}}^u_m}){e^{ - jk_c\left\| {{{\bf{r}}^u_m} - {{\bf{s}}_{1,1}}} \right\|}},{A_{1,2}}({{\bf{r}}^u_m}){e^{ - jk_c\left\| {{{\bf{r}}^u_m} - {{\bf{s}}_{1,2}}} \right\|}}}, \right.\\
	&{\left. { \cdots ,{A_{{N_d},{N_e}}}({{\bf{r}}^u_m}){e^{ - jk_c\left\| {{{\bf{r}}^u_m} - {{\bf{s}}_{{N_d},{N_e}}}} \right\|}}} \right]^H} \in \mathbb{C}^{N\times1},
\end{split}
\end{align}
where $k_c=2\pi/\lambda$ is the wave number, $\lambda$ is the wavelength, 
\begin{align}
	A_{i,l}({{\bf{r}}^u_m})=\frac{\lambda}{4\pi {{\left\| {{{\bf{r}}^u_m} - {{\bf{s}}_{i,l}}} \right\|}}}
\end{align}
denotes the path loss between the $m$th legitimate user and the $(i,l)$-th antenna\footnote{In this paper, we use $(i,l)$ to express the $l$th element of the $i$th row.} of the BS.
The received signal at the $k$th slot of the $m$th legitimate user is 
\begin{align}
	{y^u_m}(k) = {{\bf{h}}^H}({{\bf{r}}^u_m})\left(\sum\limits_{p = 1}^M {{\bf{F}}{{\bf{w}}_p}(k){x_p}(k)}\right) + {n}^u_m(k),
\end{align}
where ${x_p}(k) \in \mathbb{C}$ is the normalized data symbol intended for the $p$th legitimate user, satisfying $\mathbb{E}\left[ |{x_p(k)}|^2 \right] = 1$, ${{\bf{w}}_p}(k)\in \mathbb{C}^{N_\text{RF} \times 1}$ denotes the digital precoder for the $p$th legitimate user at the $k$th slot, ${\bf{F}} \in \mathbb{C}^{N \times N_\text{RF}}$ denotes the analog precoder matrix, whose $(n,n_\text{RF})$-th entry ${\bf{F}}_{n,n_\text{RF}}$ satisfying
\begin{align}
	\left|{\bf{F}}_{n,n_\text{RF}}\right|=1, \forall n \in \{1,2,...,N\}, \forall n_\text{RF}\in \{1,2,...,N_\text{RF}\},
\end{align}
${n}^u_m(k) \sim \mathcal{CN}(0,\sigma^2)$ denotes the additive white Gaussian noise at the $m$th legitimate user, and $\sigma^2$ is the noise power.

Without loss of generality, the total received signals at the $k$th slot of all legitimate users can be expressed in a compact form:
\begin{align}
	{\bf{y}}^u(k) = {\bf{H}}_U^H{\bf{F}}{\bf{W}}(k){\bf{x}}(k) + {\bf{n}}^u(k),
\end{align}
where 
\begin{align}
	{\bf{y}}^u(k)& ={\left[ {{y^u_1(k)},{y^u_2(k)},...,{y^u_M(k)}} \right]^\top}\in \mathbb{C}^{M \times 1},\\ 
	{\bf{x}}(k)&={\left[ {{x_1(k)},{x_2(k)},...,{x_M(k)}} \right]^\top}\in \mathbb{C}^{M \times 1},\\
	 {\bf{W}}(k) &= \left[ {{{\bf{w}}_1}(k),{{\bf{w}}_2}(k),...,{{\bf{w}}_M}}(k) \right]\in \mathbb{C}^{N_\text{RF} \times M},\\ 
	 {\bf{H}}_U &= \left[ {{\bf{h}}}({\bf{r}}^u_1),{{\bf{h}}}({\bf{r}}^u_2),...,{{\bf{h}}}({\bf{r}}^u_M) \right]\in \mathbb{C}^{N \times M}, \\
	{\bf{n}}^u(k)& ={\left[ {{n^u_1(k)},{n^u_2(k)},...,{n^u_M(k)}} \right]^\top}\in \mathbb{C}^{M \times 1}.
\end{align}

Furthermore, we assume that there are $Q$ eavesdroppers located at ${\bf{r}}^e_q = (x^e_q, y^e_q, z^e_q)$, $\forall q\in\{1,2,...,Q\}$, and the channel spanning from the $q$th eavesdropper to the BS is expressed as
\begin{align}
	\begin{split}
		{{\bf{h}}}({\bf{r}}^e_q) &= \left[ {{A_{1,1}}({{\bf{r}}^e_q}){e^{ - jk_c\left\| {{{\bf{r}}^e_q} - {{\bf{s}}_{1,1}}} \right\|}},{A_{1,2}}({{\bf{r}}^e_q}){e^{ - jk_c\left\| {{{\bf{r}}^e_q} - {{\bf{s}}_{1,2}}} \right\|}}}, \right.\\
		&{\left. { \cdots ,{A_{{N_d},{N_e}}}({{\bf{r}}^e_q}){e^{ - jk_c\left\| {{{\bf{r}}^e_q} - {{\bf{s}}_{{N_d},{N_e}}}} \right\|}}} \right]^H},
	\end{split}
\end{align}
where 
\begin{align}
	A_{i,l}({{\bf{r}}^e_q})=\frac{\lambda}{ 4\pi {{\left\| {{{\bf{r}}^e_q} - {{\bf{s}}_{i,l}}} \right\|}} }
\end{align}
denotes the path loss between the $q$th eavesdropper and the $(i,l)$-th antenna.

The received signal at the $k$th slot of the $q$th eavesdropper is
\begin{align}
	{y^e_q}(k) = {{\bf{h}}^H}({{\bf{r}}^e_q})\left(\sum\limits_{p = 1}^M {{\bf{F}}{{\bf{w}}_p}(k){x_p}(k)} \right) + {n}^e_q(k),
\end{align}
where ${n}^e_q(k) \sim \mathcal{CN}(0,\sigma^2)$ denotes the additive white Gaussian noise at the $q$th eavesdropper.

\emph{Remark 1}: In this paper, we consider a more general scenario where the non-cooperative eavesdroppers are distributed at arbitrary locations. In particular, ${{\bf{h}}}({\bf{r}}^u_m)$, $\forall m\in\{1,2,...,M\}$, are known at the BS, whereas ${{\bf{h}}}({\bf{r}}^e_q)$, $\forall q\in\{1,2,...,Q\}$, are not accessible. Thus the conventional CSI-based security beamforming techniques, e.g., \cite{6781609,7070667,9913501,9737364}, fail to work.

\section{The Precoder Design for Secure Communication}
In this section, we first introduce a low-complexity analog precoder design method based on singular value decomposition (SVD). Then, we develop a baseband SLP by adding AN to distort the signals at eavesdroppers, while enhancing the signal power at legitimate users. Moreover, we derive a time-agnostic solution that satisfies the transmit power constraint. Finally, we analyze the received signals at eavesdroppers to verify the effectiveness of the proposed algorithm and extend it to a multi-path propagation scenario.
\subsection{Low-Complexity Analog Precoder Design}
The analog precoder can be efficiently designed by utilizing the ordered SVD of the channel matrix \cite{8438554}. Specifically, the conjugate transpose of the channel matrix of all legitimate users ${\bf{H}}_U^H$ is decomposed as
\begin{align}\label{svd}
	{\bf{H}}_U^H = {\bf{C}}{\bf{\Pi }}{{\bf{D}}^H},
\end{align}
where ${\bf{C}}\in\mathbb{C}^{M \times M}$ and ${\bf{D}}\in\mathbb{C}^{N \times N}$ are unitary matrices composed of the left and right singular vectors of the channel matrix, respectively, and ${\bf{\Pi}}\in\mathbb{R}^{M \times N}$ is a matrix whose diagonal is arranged in descending order of singular values.

Let
\begin{align}
	{\bf{D}^\text{RF}} = \left[ {{\bf{d}}_1,{{\bf{d}}_{2}},...,{{\bf{d}}_{N_\text{RF}}}} \right]\in \mathbb{C}^{N \times {N_\text{RF}}},
\end{align}
where ${\bf{d}}_n\in \mathbb{C}^{N \times 1}$ denotes the $n$th column of matrix ${\bf{D}}$, $\forall n\in \{1,2,...,N_\text{RF}\}$. 
${\bf{D}^\text{RF}}$ is the matrix formed by the right singular vectors corresponding to the largest $N_\text{RF}$ singular values.
Thus the analog precoder can be designed as
\begin{align}\label{frf}
		{\bf{F}}_{n,n_\text{RF}}=e^{\left(j\angle\left({\bf{D}}_{n,n_\text{RF}}^\text{RF}\right)\right)},
\end{align}
for all $n \in \{1,2,...,N\}, n_\text{RF}\in \{1,2,...,N_\text{RF}\}$, where ${\bf{D}}_{n,n_\text{RF}}^\text{RF}$ denotes the $(n,n_\text{RF})$-th entry of ${\bf{D}^\text{RF}}$.

By selecting these dominant components of the channel matrix, the analog precoder ensures that most of the signal energy is concentrated in the legitimate user positions, resulting in a high spectral efficiency similar to that achieved by the optimum precoder but exhibits a low computational complexity \cite{8438554}.
\subsection{AN-Aided ZF Baseband Precoder Design}
Given the analog precoding matrix ${\bf{F}}$, the equivalent baseband channel matrix can be written as \cite{6928432}:
\begin{align}
{{\bf{H}}_M} = {{\bf{F}}^H}{{\bf{H}}_U}.
\end{align}
Thus, the total received signals at the $k$th slot of all legitimate users can be rewritten as:
\begin{align}
	{\bf{y}}^u(k) = {\bf{H}}_M^H{\bf{W}}(k){\bf{x}}(k) + {\bf{n}}^u(k).
\end{align}

In the context of ELAA operating in the near-field, ZF precoding becomes particularly effective due to the channel hardening effect \cite{4599181}. At high signal-to-noise ratio (SNR) regions, ZF precoding achieves performance close to that of the optimal precoding schemes, leveraging the abundant spatial degrees of freedom available in ELAA systems. Specifically, the precoder in the $k$th slot is designed to satisfy
\begin{align}\label{zf}
	{\bf{H}}_M^H{\bf{W}}(k)={{\mathbf{B }}_M},
\end{align}
where ${{\bf{B}}_M} = \text {diag}\left([ {{\beta _1},{\beta _2},...,{\beta _M}}]^\top \right)\in\mathbb{R}^{M\times M}$ denotes the expected symbol gain of the $M$ legitimate users.

The vast subspace provided by the channel matrix allows ZF to effectively mitigate interference while also creating substantial null spaces, which can be exploited to inject AN and enhance physical layer security, making it a strong candidate for secure communications. Given ${\bf{H}}_M$ and ${\bf{B}}_M$, the solution of \eqref{zf} can be expressed as
\begin{align}\label{wan}
	{\bf{W}}(k)={{\left( {{\bf{H}}_M^ \dagger } \right)}^H}{{\bf{B}}_M} +{\bf{H}}_M^ \bot {{\bf{W}}_0}(k),
\end{align}
where ${{\bf{W}}_0}(k)\in\mathbb{C}^{N_\text{RF}\times M}$ denotes the dynamic AN at symbol level, ${\bf{H}}_M^ \bot={{\bf{I}}_{N_\text{RF}} - {{\bf{H}}_M}{\bf{H}}_M^\dagger}$ is the projection matrix, and the Moore-Penrose inverse of ${\bf{H}}_M$ is expressed as
\begin{align}
	{\bf{H}}_M^\dagger = {\left( {{\bf{H}}_M^H{{\bf{H}}_M}} \right)^{ - 1}}{\bf{H}}_M^H.
\end{align}
\subsection{Dynamic AN Design with Power Constraint}
From \eqref{wan}, we observe that dynamic precoding for secure transmission can be achieved by randomizing the AN ${{\bf{W}}_0}(k)$ at the symbol level. This approach distorts signals at unintended positions while the baseband precoder ${{\bf{W}}}(k)$ is strategically determined to strengthen signals for legitimate users and eliminate inter-user interference. Nevertheless, it is crucial that this precoding approach should satisfy the transmit power constraint:
\begin{align}\label{pow}
\left\|{{\bf{F}}} {{\bf{W}}}(k) \right\|_F^2 \le P_t.
\end{align}

According to \eqref{wan}, we have	
\begin{align}\label{wf}
	\left\|{{\bf{F}}} {{\bf{W}}}(k) \right\|_F^2 
	= \left\| {\bf{F}}{{{\left( {{\bf{H}}_M^ \dagger } \right)}^H}{{\bf{B}}_M} + {\bf{F}}{\bf{H}}_M^ \bot {{\bf{W}}_0}(k)} \right\|_F^2.
\end{align}
To simplify the design of the AN matrix ${\bf{W}}_0(k)$ with power constraint \eqref{wf}, we assume that the AN has constant modulus. Specifically, let
\begin{align}
{{\bf{W}}_0}(k) = \xi {{\bf{W}}_\text{AN}}(k).
\end{align}
 Here, $\xi$ is a positive constant. The $(n_\text{RF},m)$-th entry of ${\bf{W}}_\text{AN}(k)$, is defined as ${w_{\text{AN},n,m}}(k)={e^{j{\varphi _{n_\text{RF},m}}(k)}}$, $\forall n_\text{RF} \in \{1, 2, ..., N_\text{RF}\}$, $ \forall m \in \{1, 2, ..., M\}$, and ${\varphi _{n_\text{RF},m}}(k)$ is the random phase for each entry. It is reasonable to assume that the random phases are mutually independent and uniformly distributed, i.e., ${\varphi _{n_\text{RF},m}}(k) \sim \mathcal{U}(0,2\pi) $ for each time slot $k$. Consequently, the Frobenius norm of ${\bf{W}}_0(k)$ can be expressed as $\left\| {{\bf{W}}_0}(k) \right\|_F=\sqrt{MN_\text{RF}}\xi$. 

Note that the selection of $\xi$ is closely related to the secrecy performance. A larger value of $\xi$ can increase the secrecy capacity, which will be analyzed in detail in Section IV-C. Hence, given the transmit power constraint in \eqref{pow}, we aim to determine the value of $\xi$ by solving the optimization problem
\begin{subequations}\label{problem1}
\begin{align}
	&\max \quad \xi \\
	&\text{ s.t.}\quad \left\| {{\bf{F}}{{\left( {{\bf{H}}_M^\dagger } \right)}^H}{{\bf{B}}_M} + \xi {\bf{FH}}_M^ \bot {{\bf{W}}_\text{AN}}(k)} \right\|_F^2 \le P_t,\\ 
	&\quad \quad \quad\xi >0,
\end{align}
\end{subequations}
while the solution of problem \eqref{problem1} is given by Proposition 1.

\emph{Proposition 1}: A necessary condition resulting in that problem \eqref{problem1} is solvable is 
\begin{align}\label{symgain}
 P_t>{\left\| {{\bf{F}}{{\left( {{\bf{H}}_M^\dagger } \right)}^H}{{\bf{B}}_M}} \right\|_F^2}.
\end{align}
Under this condition, the solution $\xi$ of \eqref{problem1} is explicitly given by
\begin{align}\label{eq27}
	\xi = \frac{{ - B + \sqrt {{B^2} - 4AC} }}{{2A}},
\end{align}
where we have
\begin{align}
	A &={\left\| {{\bf{FH}}_M^ \bot {{\bf{W}}_{{\text{AN}}}}(k)} \right\|_F^2},\label{AA}\\
	\begin{split}
	B &= \text{tr}\left( {{\left( {{\bf{F}}{{\left( {{\bf{H}}_M^\dagger } \right)}^H}{{\bf{B}}_M}} \right)}^H}{\bf{FH}}_M^ \bot {{\bf{W}}_{{\text{AN}}}}(k) \right.\\
		&\left.
		+ {{\left( {{\bf{FH}}_M^ \bot {{\bf{W}}_{{\text{AN}}}}(k)} \right)}^H}{\bf{F}}{{\left( {{\bf{H}}_M^\dagger } \right)}^H}{{\bf{B}}_M} \right)
	\end{split},\\
	C&={\left\| {{\bf{F}}{{\left( {{\bf{H}}_M^\dagger } \right)}^H}{{\bf{B}}_M}} \right\|_F^2} - P_t.\label{CC}
\end{align}

\emph{Proof}: Please refer to Appendix A. $\hfill \blacksquare$

Note that for high data transmission rates, determining the optimal $\xi$ at the symbol level results in a heavy computational burden. Hence, we scale the left-hand side of \eqref{pow} by finding a time-agnostic upper bound instead. Specifically, due to the fact that ${{\bf{H}}_M}{\bf{H}}_M^\dagger$ is a Hermitian matrix and ${{\bf{H}}_M^\dagger}{\bf{H}}_M={{\bf{I}}}_M$, we have 
\begin{align}\label{fro}
	\left\| {{\bf{H}}_M}{\bf{H}}_M^\dagger \right\|_F^2&=\text{tr}\left({{\bf{H}}_M}{\bf{H}}_M^\dagger\left({{\bf{H}}_M}{\bf{H}}_M^\dagger\right)^H\right)\nonumber\\
	&=\text{tr}\left({{\bf{H}}_M}{\bf{H}}_M^\dagger\right)\nonumber\\
	&=\text{tr}\left({\bf{H}}_M^\dagger{{\bf{H}}_M}\right)\nonumber\\
	&={ M }.
\end{align}
According to the triangle inequality and \eqref{fro}, we arrive at
\begin{align}
&\left\|{\bf{F}} {{\bf{H}}_M^ \bot }{{\bf{W}}_0}(k) \right\|_F\nonumber\\
\le&\sqrt{NN_\text{RF}}\left(\left\| {{\bf{W}}_0}(k) \right\|_F+\left\| {{\bf{H}}_M}{\bf{H}}_M^\dagger \right\|_F\left\| {{\bf{W}}_0}(k) \right\|_F\right)\nonumber\\
= &\sqrt{NN_\text{RF}}(1+\sqrt{ M })\left\| {{\bf{W}}_0}(k) \right\|_F. 
\end{align}
Therefore, \eqref{wf} can be further simplified as
\begin{align}\label{eq32}
	&\left\| {\bf{F}}{{{\left( {{\bf{H}}_M^ \dagger } \right)}^H}{{\bf{B}}_M} + {\bf{F}}{\bf{H}}_M^ \bot {{\bf{W}}_0}(k)} \right\|_F \nonumber\\
	\le& {{{\left\| {{\bf{F}}{{\left( {{\bf{H}}_M^\dagger } \right)}^H}{{\bf{B}}_M}} \right\|}_F} + {{\left\| {{\bf{FH}}_M^ \bot {{\bf{W}}_0}(k)} \right\|}_F}} \nonumber\\
	\le& {{{\left\| {{\bf{F}}{{\left( {{\bf{H}}_M^\dagger } \right)}^H}{{\bf{B}}_M}} \right\|}_F} + (1 + \sqrt M ){N_{{\text{RF}}}}\sqrt {MN} \xi }.
\end{align}

Based on \eqref{pow} and \eqref{eq32}, the value of $\xi$ can be selected as
\begin{align}\label{xi}
	\xi = \frac{\sqrt P_t -{\left\| {{\bf{F}}{{\left( {{\bf{H}}_M^\dagger } \right)}^H}{{\bf{B}}_M}} \right\|_F} }{(1+\sqrt{ M })N_\text{RF}\sqrt {MN}}, 
\end{align}
Note that in contrast to \eqref{eq27}, $\xi$ in \eqref{xi} is independent of the time slot, which significantly reduces the computational complexity.

As a result, the proposed precoder design algorithm is summarized as Algorithm 1.
\begin{algorithm}[t]
	\caption{Dynamic Precoder Design Algorithm}
	\label{Algorithm2}
	{\begin{tabular}{l l}
			\textbf{Input:} $ {{\bf{H}}_M} $, $P_t$, ${\bf{B}}_M$, $N_\text{RF}$ and the total time scale $K$\\
			\textbf{Output:} ${\bf{W}}(k), k=1,2,\cdots,K$. \\
			1: Obtain the analog precoder by applying \eqref{svd} -- \eqref{frf};\\
			2: \textbf{for} $k=1,2,\cdots,K$ \textbf{do}\\
			3: \ \ \quad Calculate $\xi$ according to \eqref{xi};\\
			4: \ \ \quad Randomly generate $N_\text{RF}M$ independent ${\varphi _{n_\text{RF},m}}(k)$, \\
		\ \ \quad	\quad which obey ${\varphi _{n_\text{RF},m}}(k) \sim \mathcal{U}\left( {0,2\pi } \right)$; \\
		 5: 	\ \ \quad Calculate ${{\bf{W}}_0}(k)$, whose $(n_\text{RF},m)$-th entry is given\\
		\ \ \quad	\quad by ${w_{0,n,m}}(k) = \xi{e^{j{\varphi _{n_\text{RF},m}}(k)}}, \forall n \in \{1, 2, ..., N_\text{RF}\}, $ \\
			\ \ \quad	\quad$\forall m \in \{1, 2, ..., M\}$;\\
			6: \ \ \quad Calculate the projection matrix ${\bf{H}}_M^ \bot$;\\
			7: \ \ \quad Obtain the precoder at the $k$th slot by applying \eqref{wan};\\
			8: \textbf{end} \\
	\end{tabular}}
\end{algorithm}
Note that in Algorithm 1, the secure NFC is realized by distorting the signals for passive eavesdroppers at any undesired locations. The main step of the proposed algorithm involves the SVD of the channel matrix and the calculation of ${\bf{H}}_M^\bot$, and their complexities are $O(N_\text{RF}^2M+N_\text{RF}^2N)$ and $O(N_\text{RF}^2M)$, respectively. Hence the computational complexity of the proposed algorithm is $O(N_\text{RF}^2M+N_\text{RF}^2N)$. In \cite{10436390}, the secure NFC is realized by maximizing the security capacity, which depends on the eavesdroppers' CSI, leading to a computational complexity of $O\left(l_1(M^3+N_\text{RF}^3+(l_B+1)N^3)+l_2M^3\right)$, where $l_1$, $l_B$ and $l_2$ denote the number of iterations of the block coordinate descent (BCD) loop, the bisection algorithm, and the alternating optimization (AO) loop. In general, we have $M \le N_\text{RF} \le N$ and $l_1 \gg 1$, $l_2 \gg 1$, $l_B \gg 1$. Thus, the proposed algorithm can realize secure NFC with a much lower computational complexity.
\subsection{Received Signals at Passive Eavesdroppers}
By applying the proposed precoding scheme, the signal received at the $k$th slot of the $q$th eavesdropper is
\begin{align}\label{ye}
	{y^e_q}(k) &= \sum\limits_{p = 1}^M {{{\bf{h}}^H}({\bf{r}}_q^e){\bf{F}}{{\left( {{\bf{H}}_M^\dagger } \right)}^H}{{\bf{B}}_M}{{\bf{g}}_p}{x_p}(k)} \nonumber \\
	&+\sum\limits_{p = 1}^M {{{\bf{h}}^H}({\bf{r}}_q^e){\bf{F}}{\bf{H}}_M^ \bot {{\bf{W}}_0}(k){{\bf{g}}_p}{x_p}(k)} + n_q^e(k),
\end{align} 
where ${{\bf{g}}_p}\in\left\{0,1\right\}^{M\times1}$, $p\in\{1,2,...,M\}$ is a vector with only the $p$th entry being one and other elements being zero. It is evident that at each time slot $k$, secure transmission is ensured when ${{\bf{h}}}({{\bf{r}}^e_q}) \not\in \text{span}( {{\bf{h}}}({{\bf{r}}^u_1}), \dots,{{\bf{h}}}({{\bf{r}}^u_M})), \forall q$. Specifically, if ${\bf{h}}({\bf{r}}_q^e)$ does not lie in the null subspace of the equivalent baseband channel matrix, we have ${\bf{H}}_M^\bot{\bf{F}}^H {\bf{h}}({\bf{r}}_q^e) \neq {\bf{0}}$. Consequently, the randomized ${{\bf{W}}_0}(k)$, as specified in the second term of \eqref{ye}, actively distorts the signal received by the passive eavesdroppers at each time slot. This dynamic adjustment facilitates secure communication, even in the absence of passive eavesdroppers' CSI at the BS.

\emph{Remark 3}: 
The dynamic SLP is strategically designed to enhance the signal power at legitimate users and generate random constellations at undesired positions. In contrast, conventional static precoding methods maintain unchanged constellations at undesired positions across multiple time slots. Hence, eavesdroppers can potentially decode messages successfully over time by aggregating and analyzing historical signals to learn the symbol mappings. Dynamic SLP thwarts such security threats by ensuring that the constellations are not predictable by unauthorized eavesdroppers.

\subsection{Multi-path Near-field Scenario}
Note that in practice, the wireless signal may experience multi-path propagation in the NFC scenario. The multi-path channel of the $m$th user can be modeled as \cite{10220205}:
\begin{align}\label{nlchannel}
	{\bf{h}}({\bf{r}}^u_m) = {\bf{\bar h}}({\bf{r}}^u_m) + \sum\limits_{l = 1}^L {{\alpha _l}{h_l}({{\bf{r}}^s_l},{\bf{r}}^u_m){\bf{h}}({{\bf{r}}^s_l})}, 
\end{align}
where ${\bf{\bar h}}({\bf{r}}^u_m)$ is the line-of-sight (LoS) component, ${\bf{h}}({{\bf{r}}^s_l})$ is the channel vector between the $l$th scatterer and the BS, and ${h_l}({{\bf{r}}^s_l},{\bf{r}}^u_m)$ is the channel coefficient between the $m$th user and the $l$th scatterer. $\left\{\alpha _l\right\}_{l=1}^L$ are the reflection coefficients of the scatterers, generally modeled as independently complex Gaussian distributed variables with zero mean and variance $\left\{\tilde\sigma _l^2\right\}_{l=1}^L$ \cite{10220205}.

Since ${\alpha _l}\sim \mathcal{CN}(0,\tilde\sigma _l^2)$ is independent to each other,
${{\bf{h}}}({\bf{r}}^u_m)$ is a complex Gaussian random variable. Specifically, we have ${{\bf{h}}}({\bf{r}}^u_m) \sim \mathcal{CN}({\bf{\bar h}}({\bf{r}}^u_m),{\bf{R}})$, where
\begin{align}
	{\bf{R}}&=\mathbb{E}[({{\bf{h}}}({\bf{r}}^u_m)-{\bf{\bar h}}({\bf{r}}^u_m))({{\bf{h}}}({\bf{r}}^u_m)-{\bf{\bar h}}({\bf{r}}^u_m))^H]\nonumber\\
	&=\sum\limits_{l = 1}^L {\tilde\sigma _l^2{{\left| {{h_l}({{\bf{r}}^s_l},{\bf{r}}^u_m)} \right|}^2}{\bf{h}}({{\bf{r}}^s_l})} {{\bf{h}}^H}({{\bf{r}}^s_l})
\end{align}
is the covariance matrix. It is worth noticing that once the CSI of legitimate users can be accurately estimated by the BS\footnote{Various efficient near-field channel estimation techniques have been proposed, e.g., \cite{9693928,10044679}.}, the proposed algorithm can be directly extended to the more general multi-path near-field channels.

\section{Performance Analysis}
In this section, we analyze the security performance of the proposed algorithm in terms of average SINR, achievable rate, secrecy capacity, secrecy outage probability, and secrecy zone.
\subsection{Average SINR}
The instantaneous SINR at position ${\bf{r}}$ in decoding the $m$-th information stream at any given slot $k$ can be expressed as
\begin{align}
	\text{SINR}({\bf{r}},m,k)
	&= \frac{{{{\left| {{{\bf{h}}^H}({\bf{r}}){\bf{F}}{{\bf{w}}_m}(k)} \right|}^2}}}{{\sum\limits_{p \ne m}^M {{{\left| {{{\bf{h}}^H}({\bf{r}}){\bf{F}}{{\bf{w}}_p}(k)} \right|}^2}} +\left| {n({\bf{r}},k)} \right|^2}}\nonumber\\
	&= \frac{{{{\left| {{\bf{h}}^H({\bf{r}}){\bf{F}}{\bf{W}}(k){{\bf{g}}_m}} \right|}^2}}}{{\sum\limits_{p \ne m}^M {{{\left| {{\bf{h}}^H({\bf{r}}){\bf{F}}{\bf{W}}(k){{\bf{g}}_p}} \right|}^2}} +\left| {n({\bf{r}},k)} \right|^2}}\label{sinr},
\end{align}
where ${n({\bf{r}},k)}$ is the additive white Gaussian noise at slot $k$ and position ${\bf{r}}$. Furthermore, the average SINR with respect to time slot $k$ and position ${\bf{r}}$ in decoding the $m$-th information stream can be approximated as
\begin{align}\label{sinravg}
	&\overline{\text{SINR}}({\bf{r}},m)\nonumber\\
&\simeq\frac{{\mathbb{E}\left[ {{{\bf{h}}^H}({\bf{r}}){\bf{F}}{\bf{W}}(k){\bf{g}}_m{\bf{g}}_m^H{{\bf{W}}^H}(k){\bf{F}}^H{\bf{h}}({\bf{r}})} \right]}}{{\sum\limits_{p \ne m}^M {{\mathbb{E}\left[ {{{\bf{h}}^H}({\bf{r}}){\bf{F}}{\bf{W}}(k){\bf{g}}_p{\bf{g}}_p^H{{\bf{W}}^H}(k){\bf{F}}^H{\bf{h}}({\bf{r}})} \right]}} + \sigma^2}}.
\end{align}

To analytically calculate the average SINR, the following proposition is given:
 
\emph{Proposition 2}: For any $m\in\{1,2,...,M\}$, the mathematical expectation of ${\bf{W}}(k){\bf{g}}_m{\bf{g}}_m^H{{\bf{W}}^H}(k)$ is
\begin{align}
&	\mathbb{E}\left[{\bf{W}}(k){\bf{g}}_m{\bf{g}}_m^H{{\bf{W}}^H}(k)\right] \nonumber\\
	&=	{{\xi^2} {\bf{H}}_M^\bot } 
	 + ({{\bf{H}}^\dagger_M})^H{{\bf{B }}_M}{\bf{g}}_m{\bf{g}}_m^H{\bf{B }}_M^H{{\bf{H}}^\dagger_M}.
\end{align}

\emph{Proof}: Please refer to Appendix B. $\hfill \blacksquare$

Assuming that in a multi-path propagation scenario, the CSI at an unintended location ${\bf{r}}$ is independent of the legitimate users, and therefore independent of the precoding matrix ${\bf{W}}(k)$. Additionally, the CSI of the legitimate users is perfectly known at the BS. By using Proposition 2, notice that $\mathbb{E}\left( {{\text{tr}}\left( \cdot \right)} \right) = {\text{tr}}\left( {\mathbb{E}\left( \cdot \right)} \right)$ and ${\text{tr}}\left({\bf{X}}{\bf{Y}} \right)={\text{tr}}\left({\bf{Y}}{\bf{X}} \right)$, the mathematical expectation of ${{\bf{h}}^H}({\bf{r}}){\bf{F}}{\bf{W}}(k){\bf{g}}_m{\bf{g}}_m^H{{\bf{W}}^H}(k){\bf{F}}^H{\bf{h}}({\bf{r}})$ can be calculated as
\begin{align}\label{sinrexp}
	&\mathbb{E}\left[{{\bf{h}}^H}({\bf{r}}){\bf{F}}{\bf{W}}(k){\bf{g}}_m{\bf{g}}_m^H{{\bf{W}}^H}(k){\bf{F}}^H{\bf{h}}({\bf{r}})\right] \nonumber\\
	\begin{split}
		=& {\text{tr}}\left( \mathbb{E}\left[{\bf{h}}({\bf{r}}){{\bf{h}}^H}({\bf{r}})\right] \right.\\
		&\left.
		{\bf{F}}\left({{\xi^2} {\bf{H}}_M^\bot }
		+ ({{\bf{H}}^\dagger_M})^H{{\bf{B }}_M}{\bf{g}}_m{\bf{g}}_m^H{\bf{B }}_M^H{{\bf{H}}^\dagger_M} \right){\bf{F}}^H \right).
	\end{split}
\end{align}

In the LoS propagation scenario, ${{\bf{h}}}({\bf{r}})$ is deterministic. Substituting \eqref{sinrexp} into \eqref{sinravg}, the average SINR can be approximated by \eqref{sinrlos}. Specifically, when ${\bf{r}}={\bf{r}}^u_m$, notice that ${\bf{H}}_M^\dagger {\bf{F}}^H{\bf{h}}({\bf{r}}^u_m)={\bf{g}}_m$ and ${\bf{H}}_M{\bf{g}}_m={\bf{F}}^H{\bf{h}}({\bf{r}}^u_m)$, we can achieve $\overline{\text{SINR}}({\bf{r}}^u_m,m)\simeq{\beta _m^2}/{\sigma^2}$.

On the other hand, in a multi-path propagation scenario, ${{\bf{h}}}({\bf{r}}) \sim \mathcal{CN}({\bf{\bar h}}({\bf{r}}),{\bf{R}})$ is a complex Gaussian random variable, $\mathbb{E}\left[{\bf{h}}({\bf{r}}){{\bf{h}}^H}({\bf{r}})\right]={{\bf{R}} + {\bf{\bar h}}({\bf{r}}){{{\bf{\bar h}}}^H}({\bf{r}})}$. Similarly, the average SINR can be approximated as shown in \eqref{sinrnlos}.
\begin{figure*}
	\begin{align}\label{sinrlos}
		\overline{\text{SINR}}({\bf{r}},m)_{{\text{LoS}}}
		\simeq \frac{{
						{{\xi^2} {{\bf{h}}^H}({\bf{r}}){\bf{F}}{\bf{H}}_M^\bot{\bf{F}}^H {\bf{h}}({\bf{r}})} 
						{ + {{\left| {{{\bf{h}}^H}({\bf{r}}){\bf{F}}{{({\bf{H}}_M^\dagger )}^H}{{\bf{B}}_M}{{\bf{g}}_m}} \right|}^2}}
				 }}{{
						{(M - 1){\xi^2}{{\bf{h}}^H}({\bf{r}}){\bf{F}}{\bf{H}}_M^\bot{\bf{F}}^H {\bf{h}}({\bf{r}})}
						{ + \sum\limits_{p \ne m}^M {{{\left| {{{\bf{h}}^H}({\bf{r}}){\bf{F}}{{({\bf{H}}_M^\dagger )}^H}{{\bf{B}}_M}{{\bf{g}}_p}} \right|}^2}} }
				 + \sigma^2}}.
	\end{align}
	\begin{align}\label{sinrnlos}
		\overline{\text{SINR}}({\bf{r}},m)_{{\text{MP}}}\simeq\frac{{{\text{tr}}\left( {\left( {{\bf{R}} + {\bf{\bar h}}({\bf{r}}){{{\bf{\bar h}}}^H}({\bf{r}})} \right){\bf{F}}\left( 
							{{\xi^2}{\bf{H}}_M^\bot}
							+{({{\bf{H}}^\dagger_M})^H{{\bf{B }}_M}{\bf{g}}_m{\bf{g}}_m^H{\bf{B }}_M^H{{\bf{H}}^\dagger_M}}
					 \right){\bf{F}}^H} \right)}}{{\sum\limits_{p \ne m}^M {{\text{tr}}\left( {\left( {{\bf{R}} + {\bf{\bar h}}({\bf{r}}){{{\bf{\bar h}}}^H}({\bf{r}})} \right){\bf{F}}\left( 
								{{\xi^2}{\bf{H}}_M^\bot}
								{ + ({{\bf{H}}^\dagger_M})^H{{\bf{B }}_M}{\bf{g}}_p{\bf{g}}_p^H{\bf{B }}_M^H{{\bf{H}}^\dagger_M}} \right)}{\bf{F}}^H \right)} + \sigma^2}}.
	\end{align}\hrulefill
\end{figure*}

\subsection{Achievable Rate and Secrecy Capacity}
By exploiting the concave property of function $\log_2(1+x)$ and Jensen's inequality, the upper bound of the average achievable rate at position ${\bf{r}}$ of the $m$th user is
\begin{align}\label{upper}
	\widetilde{C}({\bf{r}},m)=\log_2 (1 + \overline{\text{SINR}}({\bf{r}},m)),
\end{align}
where $\overline{\text{SINR}}({\bf{r}},m)$ is given by \eqref{sinrlos} or \eqref{sinrnlos}, depending on the interested scenario. 

Considering fast-fading channels, the secrecy capacity between the $m$th legitimate user and the $q$th eavesdropper is defined as the mathematical expectation of the difference between the legitimate channel capacity and the eavesdropping channel capacity:
\begin{align}
	\overline{C}_s({\bf{r}}^u_m,{\bf{r}}^e_q)=\max \left\{ {\mathbb{E}\left[ {{{\log }_2}\left( {\frac{{1 + {\text{SINR}}({\bf{r}}_m^u,m,k)}}{{1 + {\text{SINR}}({\bf{r}}_q^e,m,k)}}} \right)} \right],0} \right\}.
\end{align}

In this subsection, we provide an approximation for the secrecy capacity by using the upper bound in \eqref{upper}, which provides a closed-form expression for convenient evaluation of secrecy performance:
\begin{align}\label{securitycap}
		\overline{C}_s({\bf{r}}^u_m,{\bf{r}}^e_q) \simeq \max \left\{ {\widetilde{C}({{\bf{r}}^u_m},m) - 	\widetilde{C}({{\bf{r}}^e_q},m),0} \right\},
\end{align}
where
\begin{align}
	\widetilde{C}({\bf{r}}^u_m,m)&=\log_2 \left(1 + \frac{{\beta _m^2}}{{\sigma^2}}\right),\\
	\widetilde{C}({\bf{r}}^e_q,m)&=\log_2 (1 + \overline{\text{SINR}}({\bf{r}}^e_q,m)), 
\end{align}
and $\overline{\text{SINR}}({\bf{r}}^e_q,m)$ can be found by substituting ${\bf{r}}^e_q$ into \eqref{sinrlos} or \eqref{sinrnlos}. It is worth noting that the exact secrecy capacity can be obtained by deriving the corresponding probability density function (PDF), as will be illustrated in Section IV-C. 

From \eqref{sinrlos} and \eqref{securitycap}, we observe that when the inequality
\begin{align}
	&(M-1){{{\left| {{{\bf{h}}^H}({\bf{r}}^e_q){\bf{F}}{{({\bf{H}}_M^\dag )}^H}{{\bf{B}}_M}{{\bf{g}}_m}} \right|}^2}}\nonumber\\
	>&{\sum\limits_{p \ne m}^M {{{\left| {{{\bf{h}}^H}({\bf{r}}^e_q){\bf{F}}{{({\bf{H}}_M^\dag )}^H}{{\bf{B}}_M}{{\bf{g}}_p}} \right|}^2}} + \sigma^2},
\end{align}
is satisfied, the secrecy capacity $\widetilde{C}_s({\bf{r}}^u_m, {\bf{r}}^e_q)$ in the LoS channel becomes a monotonically increasing function of $\xi$. Consequently, given a desired secrecy capacity $\delta$, the parameter $\xi$ should be chosen to satisfy \eqref{xilos}, at the top of the next page, to achieve the desired secrecy capacity.
\begin{figure*}
\begin{align}\label{xilos}
\xi \ge \sqrt {\frac{{{{\left| {{{\bf{h}}^H}({{\bf{r}}^e_q}){\bf{F}}{{({\bf{H}}_M^\dagger )}^H}{{\bf{B}}_M}{{\bf{g}}_m}} \right|}^2} - \left( {\sum\limits_{p \ne m}^M {{{\left| {{{\bf{h}}^H}({{\bf{r}}^e_q}){\bf{F}}{{({\bf{H}}_M^\dagger )}^H}{{\bf{B}}_M}{{\bf{g}}_p}} \right|}^2}} + \sigma^2} \right)\left( {{2^{ - \delta }}\left( {1 + \frac{{\beta _m^2}}{\sigma^2}} \right) - 1} \right)}}{{{{\bf{h}}^H}({{\bf{r}}^e_q}){\bf{F}}{\bf{H}}_M^ \bot {\bf{F}}^H{\bf{h}}({{\bf{r}}^e_q})\left( {\left( {M - 1} \right)\left( {{2^{ - \delta }}\left( {1 + \frac{{\beta _m^2}}{\sigma^2}} \right) - 1} \right) - 1} \right)}}}, 
\end{align}
\end{figure*}
 \begin{figure*}[!t]
	\begin{align}\label{security_hisnr}
		{f_{(M-1){\text{SINR}}({{\bf{r}}^e_q},m,k)}}(s)={\sum\limits_{k = 0}^\infty {\sum\limits_{l = 0}^\infty {\frac{{{2^{ - (k + l)}}{{(M - 1)}^{l + M - 1}}{s^k}\lambda _{nc1}^k\lambda _{nc2}^l}}{{k!l!{e^{({\lambda _{nc1}} + {\lambda _{nc2}})/2}}{\cal B}\left( {k + 1,l + M - 1} \right)}}} } {{\left( {M - 1 + s} \right)}^{ - (k + l) - M}}},s>0.\tag{62}
	\end{align}\hrulefill
\end{figure*}
In the multi-path channel, the corresponding parameter $\xi$ can be obtained similarly by using \eqref{sinrnlos}.
\subsection{Secrecy Outage Probability}
When considering slow-fading channels, the secrecy outage probability of the $m$th legitimate user under the interception of the $q$th eavesdropper is defined as the probability that the secrecy capacity falls below a predefined rate $R_s$:
\begin{align} 
&\text{Pr}_\mathrm{out}({R_s}) = \text{Pr}({C}_s({\bf{r}}^u_m,{\bf{r}}^e_q) < {R_s})\nonumber\\
&=\text{Pr}\left( {{\text{SINR}}({{\bf{r}}^e_q},m,k) > \frac{{1 + {\text{SINR}}({{\bf{r}}^u_m},m,k)}}{{{2^{{R_s}}}}} - 1} \right).
\end{align}
Next, to calculate the secrecy outage probability, we derive the distribution of ${\text{SINR}}({{\bf{r}}^u_m},m,k)$ and ${\text{SINR}}({{\bf{r}}^e_q},m,k)$. 

The expression of $\text{SINR}({\bf{r}}^u_m,m,k)$ is given by
\begin{align}
	\text{SINR}({\bf{r}}^u_m,m,k)=\frac{{\beta _m^2}}{{{{\left| {n}^u_m(k) \right|}^2}}}.
\end{align}
Considering the white Gaussian noise $ {{n}^u_m(k)} \sim \mathcal{CN}(0,\sigma^2)$, ${\left| {n}^u_m(k) \right|}^2$ follows an exponential distribution. Hence, the PDF of ${\left| {n}^u_m(k)\right|}^2$ is
\begin{align}
	&{f_{\left| {n}^u_m(k) \right|^2}}(x) =\frac{1}{\sigma^2}{e^{\frac{{ - x}}{\sigma^2}}},x>0.
\end{align}
By exploiting the Jacobi transformation, the PDF of $\text{SINR}({\bf{r}}^u_m,m,k)$ is
\begin{align}
f_{{\text{SINR}}({{\bf{r}}^u_m},m,k)}(y)=\frac{{\beta _m^2}}{{\sigma^2{y^2}}}{e^{\frac{{ - \beta _m^2}}{{\sigma^2y}}}},y>0.
\end{align}

Furthermore, we consider the worst case where the eavesdropper channel is noiseless, $n_q^e(k) = 0$, $\forall q\in\{1,2,...,Q\}$. Substituting \eqref{wan} into \eqref{sinr}, we have
\begin{align}
	&\text{SINR}({\bf{r}}^e_q,m,k)\nonumber\\
	&=\frac{{{{\left| {{{\bf{h}}^H}({{\bf{r}}^e_q}){\bf{F}}{\bf{H}}_M^ \bot {{\bf{W}}_0}(k){{\bf{g}}_m} + {{\bf{h}}^H}({{\bf{r}}^e_q}){\bf{F}}{{({\bf{H}}_M^\dagger )}^H}{{\bf{B}}_M}{{\bf{g}}_m}} \right|}^2}}}{{\sum\limits_{p \ne m}^M {{{\left| {{{\bf{h}}^H}({{\bf{r}}^e_q}){\bf{F}}{\bf{H}}_M^ \bot {{\bf{W}}_0}(k){{\bf{g}}_p} + {{\bf{h}}^H}({{\bf{r}}^e_q}){\bf{F}}{{({\bf{H}}_M^\dagger )}^H}{{\bf{B}}_M}{{\bf{g}}_p}} \right|}^2}} }}\nonumber\\
	&=\frac{{{{\left| {{{\bf{h}}^H}({{\bf{r}}^e_q}){\bf{F}}{\bf{H}}_M^ \bot {{\bf{w}}_{0,m}}(k) + {{\bf{h}}^H}({{\bf{r}}^e_q}){\bf{F}}{{({\bf{H}}_M^\dagger )}^H}{{\bf{B}}_M}{{\bf{g}}_m}} \right|}^2}}}{{\sum\limits_{p \ne m}^M {{{\left| {{{\bf{h}}^H}({{\bf{r}}^e_q}){\bf{F}}{\bf{H}}_M^ \bot {{\bf{w}}_{0,p}}(k) + {{\bf{h}}^H}({{\bf{r}}^e_q}){\bf{F}}{{({\bf{H}}_M^\dagger )}^H}{{\bf{B}}_M}{{\bf{g}}_p}} \right|}^2}} }}\nonumber\\
	&=\frac{{{{\left| {{\bf{u}}_q^H} {{\bf{w}}_{0,m}}(k) + v_{q,m} \right|}^2}}}{{\sum\limits_{p \ne m}^M {{{\left| {{\bf{u}}_q^H} {{\bf{w}}_{0,p}}(k) + v_{q,p}\right|}^2}} }},
\end{align}
where
\begin{align}
	{{\bf{u}}_q^H} &= {{\bf{h}}^H}({{\bf{r}}^e_q}){\bf{F}}{\bf{H}}_M^ \bot,\\ v_{q,m}&={{\bf{h}}^H}({{\bf{r}}^e_q}){\bf{F}}{{({\bf{H}}_M^\dagger )}^H}{{\bf{B}}_M}{{\bf{g}}_m}. 
\end{align}
Moreover, 
\begin{align}
	\left| {{\bf{u}}_q^H} {{\bf{w}}_{0,m}}(k) + v_{q,m} \right|^2={{\left| {\sum\limits_{n = 1}^N {\xi u_{q,n}^*{e^{j{\varphi _{n,m}(k)}}} + v_{q,m}} } \right|}^2},
\end{align}
where $u_{q,n}$ denotes the $n$th entry of ${\bf{u}}_q$.

In the multi-user NFC system, ELAA is equipped with a large number of antenna elements. It is important to note that $e^{j{\varphi _{n,m}(k)}}$ is independent of each other, $\forall n,m,k$. In the LoS scenario, ${{\bf{h}}}({\bf{r}}^e_q)$ is deterministic. According to the central limit theorem and Proposition 1, we can conclude that $\sum\limits_{n = 1}^N {\xi u_{q,n}^*{e^{j{\varphi _{n,m}(k)}}}} \to \mathcal{CN}(0,\xi^2\left\| {\bf{u}}_q \right\|^2)$, for $N \to \infty$. Therefore, we have 
\begin{align}
	\psi_{q,m}&=\frac{\sqrt{2}}{{\xi \left\| {\bf{u}}_q \right\|}}\left( {\sum\limits_{n = 1}^N {\xi u_{q,n}^*{e^{j{\varphi _{n,m}(k)}}}} + {v_{q,m}}} \right)\nonumber \\
	&\sim \mathcal{CN}(\frac{\sqrt{2}v_{q,m}}{{\xi \left\| {\bf{u}}_q \right\|}},2).
\end{align}
Furthermore, we can derive that $\Re\left(\psi_{q,m}\right)\sim \mathcal{N}\left(\Re\left(\frac{\sqrt{2}v_{q,m}}{{\xi \left\| {\bf{u}}_q \right\|}}\right),1\right)$, $\Im\left(\psi_{q,m}\right)\sim \mathcal{N}\left(\Im\left(\frac{\sqrt{2}v_{q,m}}{{\xi \left\| {\bf{u}}_q \right\|}}\right),1\right)$, and thus $|\psi_{q,m}|^2 =\frac{2}{{\xi^2 \left\| {\bf{u}}_q \right\|}^2}{{\left| {\sum\limits_{n = 1}^N {\xi u_{q,n}^*{e^{j{\varphi_{n,m}(k)}}} + v_{q,m}} } \right|}^2}$ follows the non-central chi-square distribution with 2 DoF, the non-centrality parameter $\lambda_{nc1}$ is
\begin{align}
	\lambda_{nc1}&=\left(\mathbb{E}\left[\Re\left(\psi_{q,m}\right)\right]\right)^2+\left(\mathbb{E}\left[\Im\left(\psi_{q,m}\right)\right]\right)^2\nonumber\\
	&=\frac{2|v_{q,m}|^2}{{\xi^2 \left\| {\bf{u}}_q \right\|}^2}.
\end{align}

Similarly, $\sum\limits_{p \ne m}^M {|{\psi _{q,p}}{|^2}}= \frac{2}{{\xi^2 \left\| {\bf{u}}_q \right\|}^2}\sum\limits_{p \ne m}^M\left| {{\bf{u}}_q^H} {{\bf{w}}_{0,p}}(k) + v_{q,p} \right|^2$ follows the non-central chi-square distribution with $2(M-1)$ DoF. The non-centrality parameter $\lambda_{nc2}$ can be expressed as
 \begin{align}
 	\lambda_{nc2}&=\sum\limits_{p \ne m}^M \left( \left(\mathbb{E}\left[\Re\left(\psi_{q,p}\right)\right]\right)^2+\left(\mathbb{E}\left[\Im\left(\psi_{q,p}\right)\right]\right)^2\right)\nonumber\\
 	&=\frac{2\sum\limits_{p \ne m}^M|v_{q,p}|^2}{{\xi^2 \left\| {\bf{u}}_q \right\|}^2}.
 \end{align}\setcounter{equation}{62}

In conclusion, ${\text{SINR}}({{\bf{r}}^e_q},m,k)$ is a quotient of two non-central chi-square variables, with their DoF and non-centrality parameter are $(2, \lambda_{nc1})$ and $(2(M-1), \lambda_{nc2})$, respectively. Consequently, $(M-1){\text{SINR}}({{\bf{r}}^e_q},m,k)$ follows a doubly non-central $F$ distribution \cite{ncf}. Its corresponding PDF can be expressed as
 \eqref{security_hisnr}, where 
 \begin{align}
 \mathcal{B}(a,b)=\int_0^1 {{x^{a - 1}}{{(1 - x)}^{b - 1}}{\rm{d}}x} 
 \end{align}
 denotes the Beta function.
 
As a result, the secrecy outage probability is given as follows:
 	\begin{align}\label{securityp}
 	&\text{Pr}_\mathrm{out}({R_s}) \nonumber\\
=&\int_0^\infty {\int_0^{{2^{{R_s}}}(\frac{s}{{M - 1}} + 1) - 1} 
	\hspace{-10.5mm}
	{{f_{(M - 1){\text{SINR}}({{\bf{r}}^e_q},m,k)}}(s){f_{{\text{SINR}}({{\bf{r}}^u_m},m,k)}}(y){\rm{d}}y{\rm{d}}s} } \nonumber\\
=&\int_0^\infty {{e^{ - \frac{{\beta _m^2}}{{{{ \sigma}^2}\left( {{2^{{R_s}}}(\frac{s}{{M - 1}} + 1) - 1} \right)}}}}{f_{(M - 1){\text{SINR}}({{\bf{r}}^e_q},m,k)}}(s)} {\rm{d}}s.
\end{align}
\begin{figure}[!t]
	\centering
	\subfloat[]{
		\label{pos1}
		\includegraphics[width=0.25\textwidth]{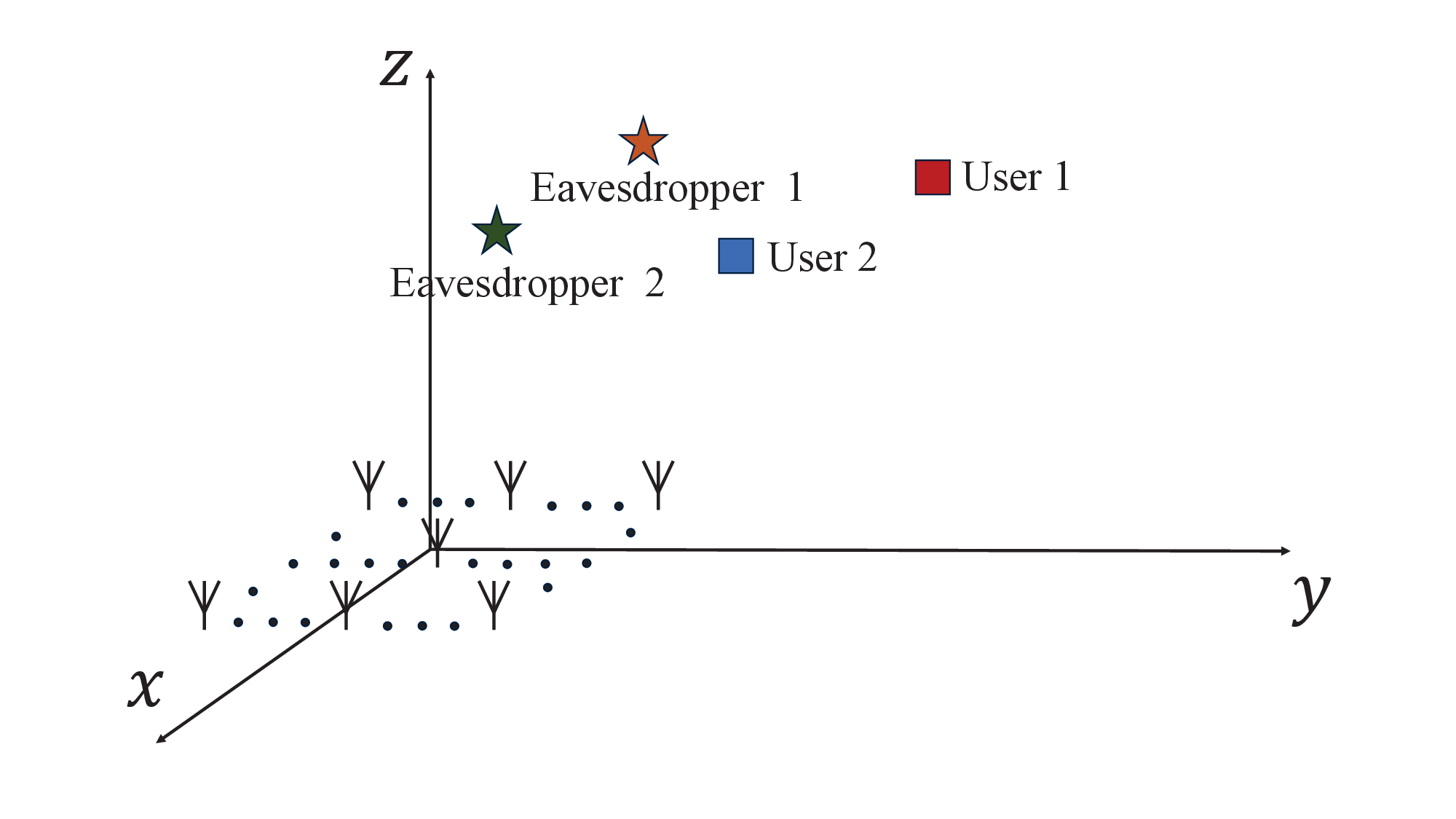}}	
	\centering
	\subfloat[]{
		\label{samepos}
		\includegraphics[width=0.25\textwidth]{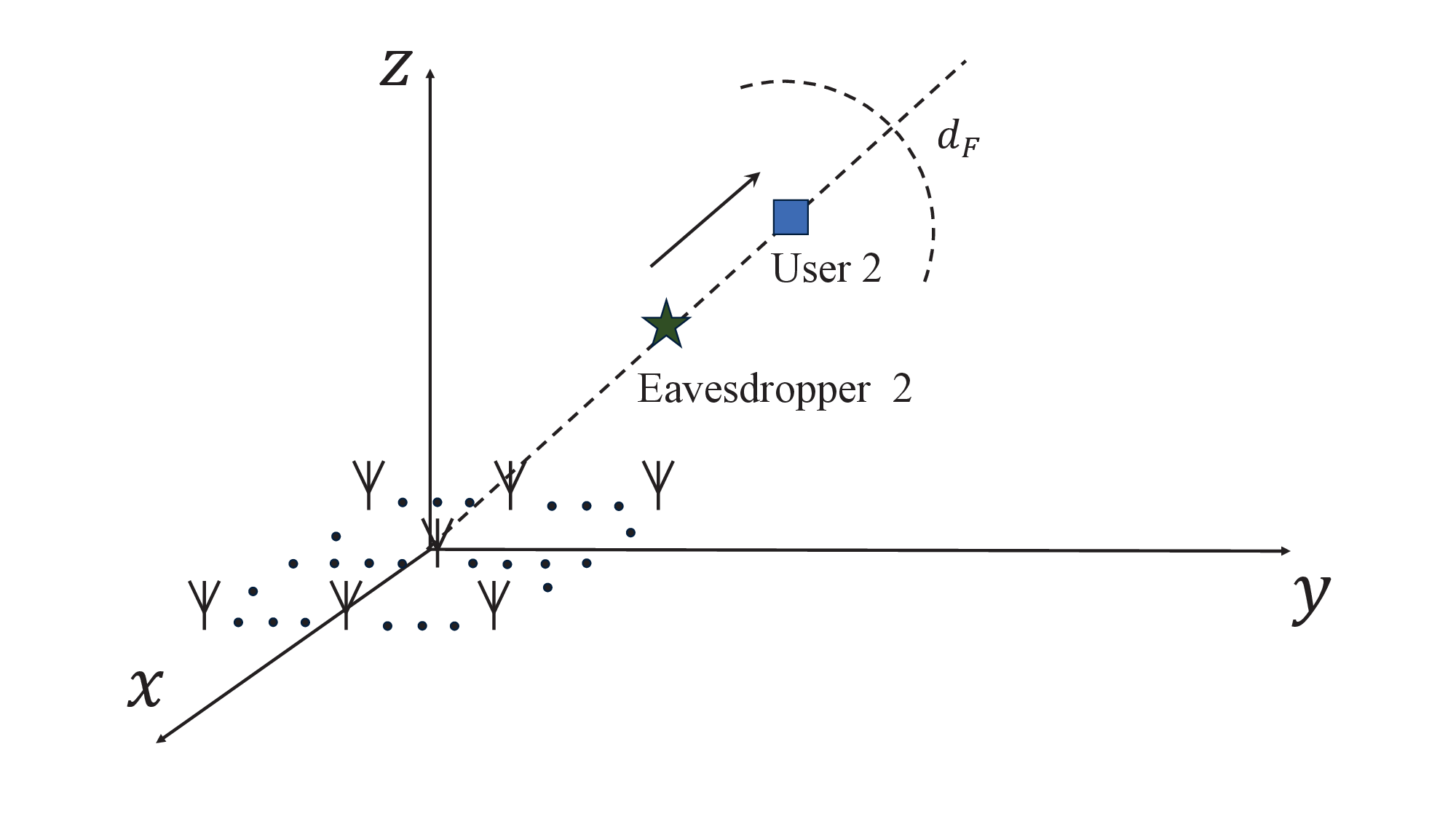}}	
	\caption{The illustration of Users and Eavesdroppers, where: (a) Fixed users and eavesdroppers positions; (b) Same direction for User 2 and Eavesdropper 2.}
		\label{pos}
\end{figure}
\subsection{Secrecy Zone}
The secrecy zone measures the confidentiality performance of a system based on the location information of users. It refers to an area where eavesdroppers are unable to intercept information of legitimate users, which is defined as the geometric area where the secrecy outage probability is lower than a given threshold value $\varepsilon$:
\begin{align}\label{zone}
	\Theta \left( {{{\bf{r}}_m}} \right) = \left\{ {{\bf{r}}|{{\Pr }_\mathrm{out}}({R_s}) \le \varepsilon } \right\}.
\end{align}

Given a fixed ${{{\bf{r}}_m}}$, the corresponding secrecy zone can be determined by substituting \eqref{securityp} into \eqref{zone}.

It should be noted that the derivation of secrecy outage probability and the secrecy zone is predicated on near-field LoS channels, owing to the independence requirement of the central limit theorem. In near-field multi-path propagation scenarios, the passive eavesdroppers' channels are unavailable at the BS, and thus ${{\bf{u}}}_q$ and $v_{q,m}$ are treated as random variables with dependent components. Consequently, determining the secrecy outage probability and the secrecy zone in the context of near-field multi-path channels poses significant challenges, which we leave for future investigation.
\section{Simulation Results}
In this section, we provide extensive simulation results to verify the security performance of the proposed algorithm. The carrier frequency is set to $f_c=28$ GHz. We adopt a $40 \times 40 $ uniform square planar array positioned in the $xy$-plane, where the spacing between the array elements is $d=0.5\lambda=5.4$ mm, resulting in the antenna length of $L=20.9$ cm. The antenna aperture is $D = \sqrt{2}L$. Consequently, the Fraunhofer distance is calculated as ${d_F} = 2{D^2}/\lambda = 16.3$ m. For the analog precoder, we set the number of RF chains to $N_\text{RF}= 40$. As illustrated in Fig. \ref{pos}\subref{pos1}, unless otherwise specified, two legitimate users, User 1 and User 2, are positioned at $(-0.3d_F,0.5d_F,0.55d_F)$ and $(0.2d_F,0.3d_F,0.55d_F)$, respectively, while Eavesdropper 1 and Eavesdropper 2 are located at $(-0.4d_F,0.2d_F,0.55d_F)$ and $(0.1d_F,0.1d_F,0.55d_F)$. It is worth noting that, although the carrier frequency used in our simulations is $f_c=28$ GHz, the simulation results are also applicable at higher frequencies, as we fix the antenna spacing at half-wavelength and normalize the positions of both the users and the eavesdropper with respect to the Fraunhofer distance. Furthermore, Eavesdropper 1 and Eavesdropper 2 attempt to decode the information streams intended for User 1 and User 2, respectively, The legitimate users' CSI is perfectly known, while the CSI of the eavesdroppers is not available at the BS. 
\begin{figure}[!t]
	\centering
	\subfloat[]{
		\label{bp1}
		\includegraphics[width=0.25\textwidth]{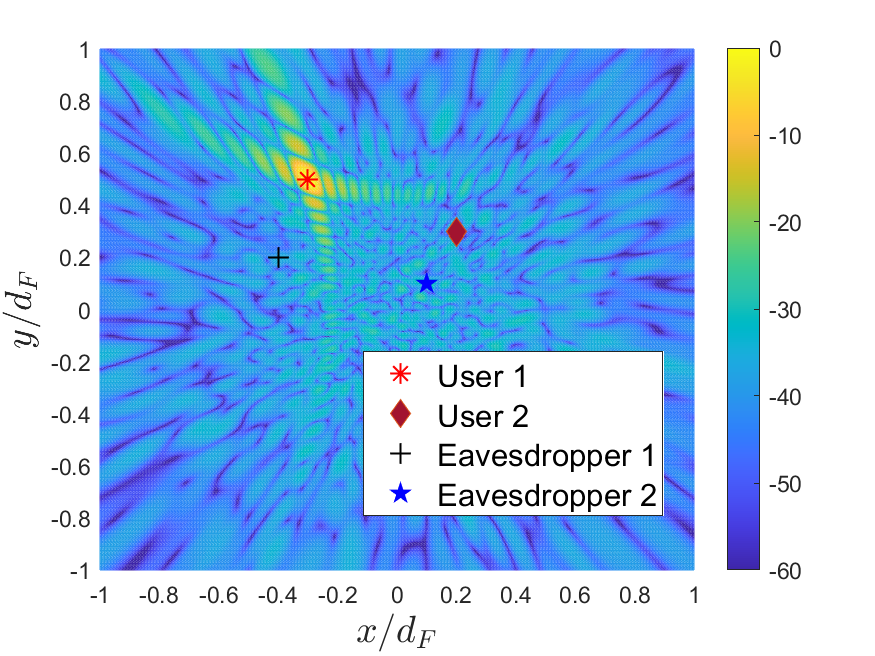}}
	\subfloat[]{
		\label{bp2}
		\includegraphics[width=0.25\textwidth]{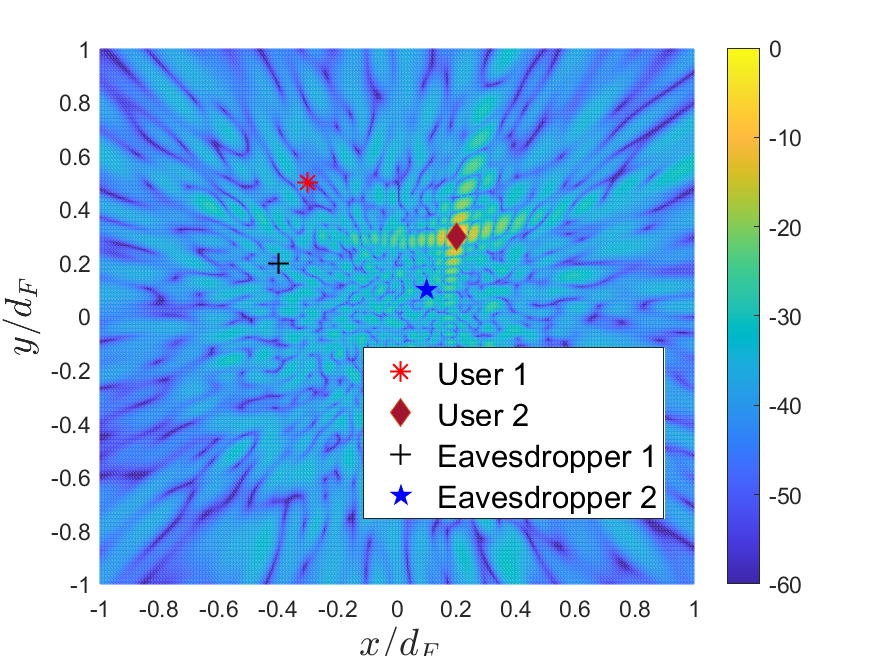}}
	\caption{The top view of beampatterns at $z=0.55d_F$, where: (a) Beampattern of the precoder for User 1; (b) Beampattern of the precoder for User 2.}
	\label{bp}
\end{figure}
\begin{figure}[!t]
	\centering
	\subfloat[]{
		\label{bp1xoz}
		\includegraphics[width=0.25\textwidth]{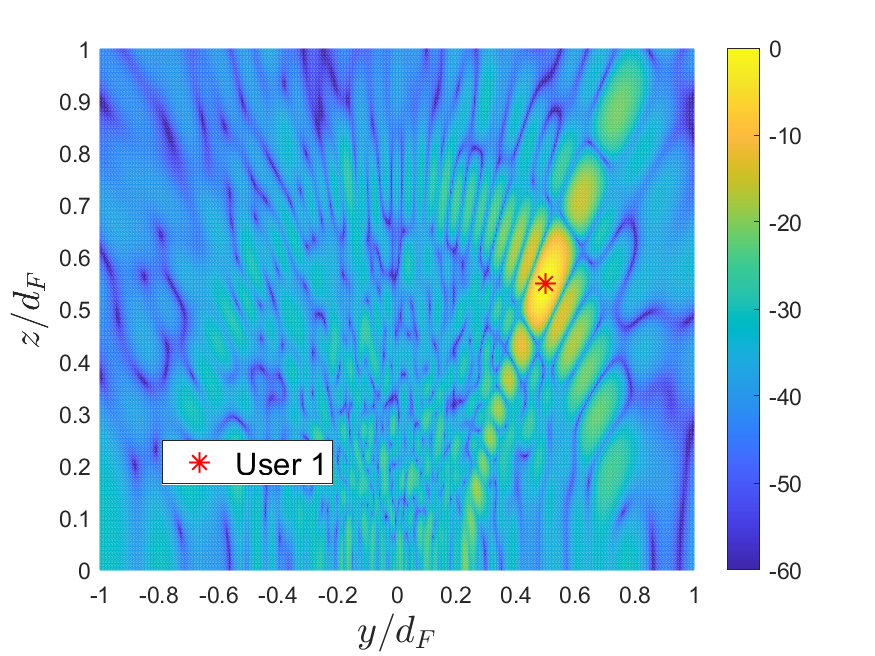}}
	\subfloat[]{
		\label{bp2xoz}
		\includegraphics[width=0.25\textwidth]{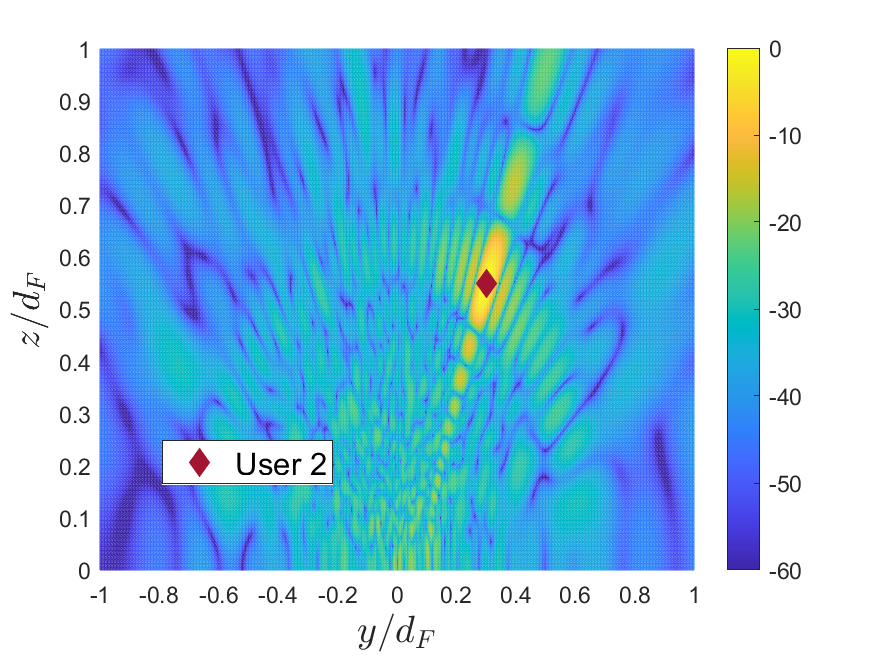}}
	\caption{The side view of beampatterns, where: (a) Beampattern of the precoder for User 1 at $x=-0.3d_F$; (b) Beampattern of the precoder for User 2 at $x=0.2d_F$.}
	\label{bpxoz}
\end{figure}
\begin{figure}[!t]
	\subfloat[]{
		\label{8psk}
		\includegraphics[width=0.25\textwidth]{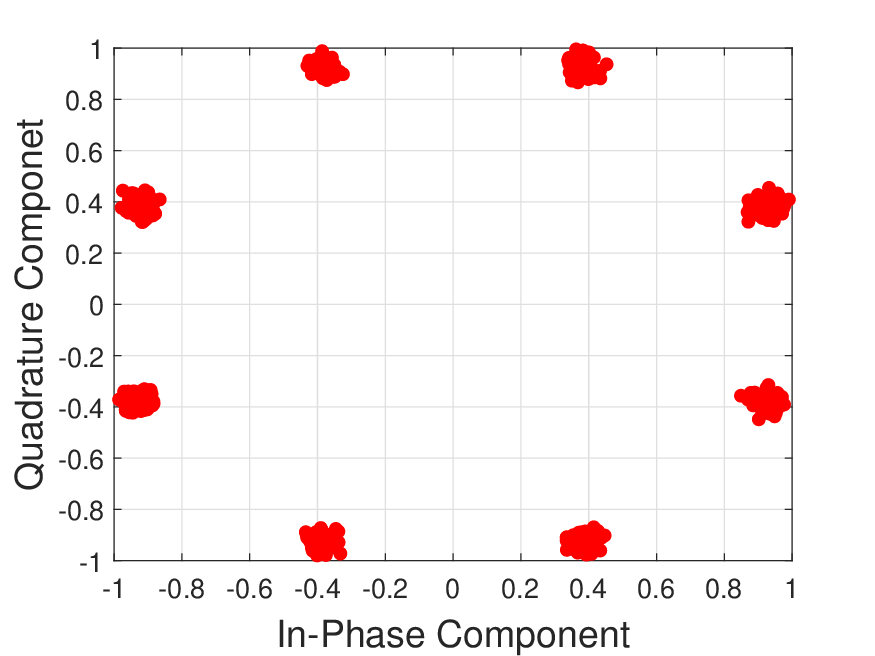}}
	\subfloat[]{
		\label{qam}
		\includegraphics[width=0.25\textwidth]{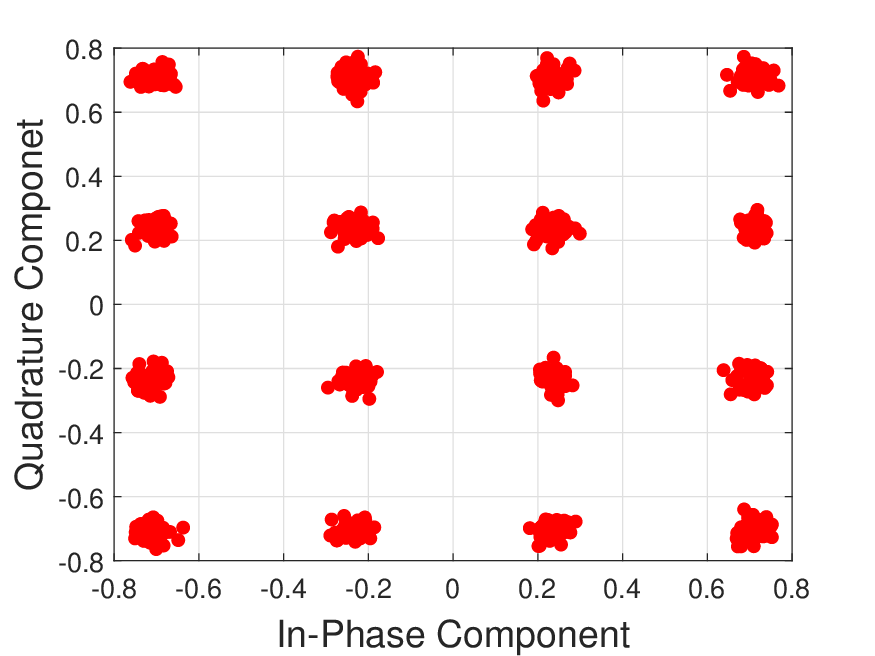}}\
	\subfloat[]{
		\label{8pskeve}
		\includegraphics[width=0.25\textwidth]{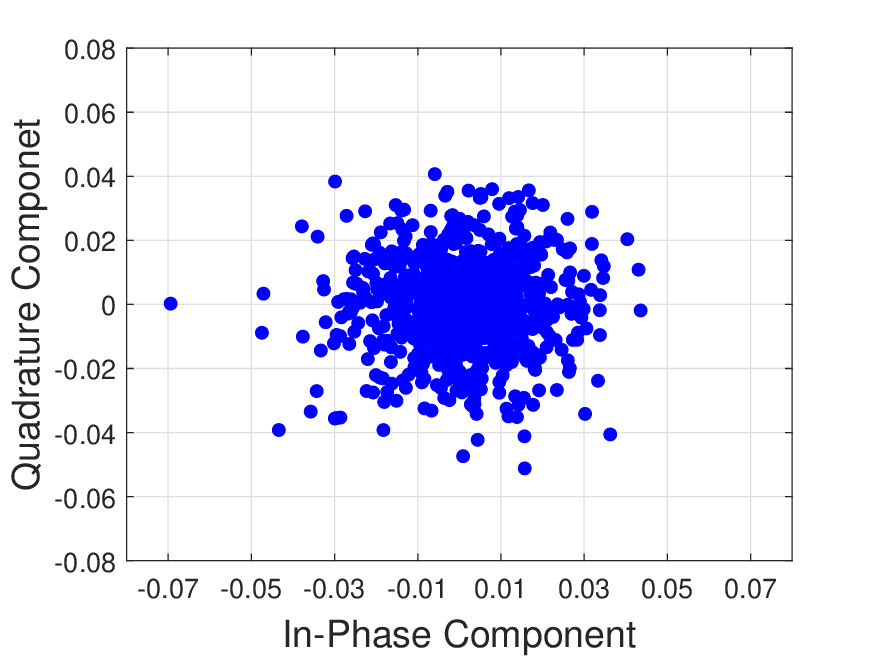}}
	\subfloat[]{
		\label{qameve}
		\includegraphics[width=0.25\textwidth]{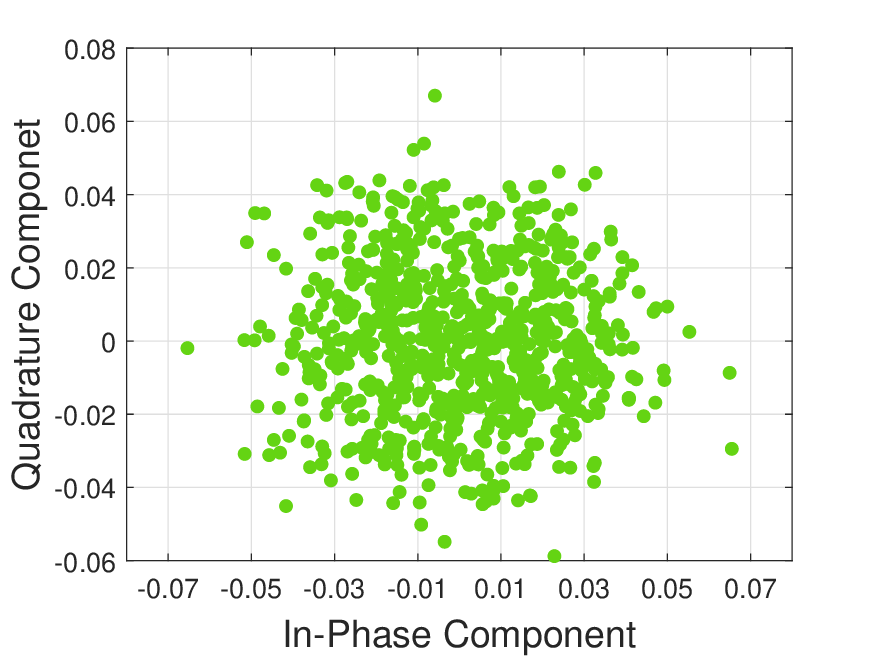}}
	\caption{Constellation diagram of received signals at legitimate users and eavesdroppers relying on the proposed algorithm, where the SINR for legitimate users is $15$ dB, and the number of total time slots is $K=800$. (a) User 1, 8-PSK; (b) User 2, 16-QAM; (c) Eavesdropper 1, chaotic constellation; (d) Eavesdropper 2, chaotic constellation.}
	\label{Constellations}
\end{figure}
\subsection{Multi-user Beamforming Gain}
In this subsection, we examine the beamforming gain by analyzing the three-dimensional near-field beam response on specific planes, specifically at $z = 0.55d_F$, $x = -0.3d_F$, and $x = 0.2d_F$. The symbol gains are uniformly set to $\beta_1=\beta_2=1$, and according to \eqref{symgain}, the transmit power is $P_t = 5{\left\| {{\bf{F}}{{\left( {{\bf{H}}_M^\dagger } \right)}^H}{{\bf{B}}_M}} \right\|_F^2}$ for illustration. The top views of the beampatterns are depicted in Fig. \ref{bp}, and the side views of the beampatterns are depicted in Fig. \ref{bpxoz}. From Figs. \ref{bp} and \ref{bpxoz}, we can see that the beams are accurately focused at the positions of legitimate users. In Fig. \ref{bp}\subref{bp1}, the beamforming gain at User 1 is exploited as a reference, with the normalized beamforming gains at User 2, Eavesdropper 1, and Eavesdropper 2 being 0, $-36.9$ dB and $-34.0$ dB, respectively. In Fig. \ref{bp}\subref{bp2}, the beamforming gain at User 2 serves as the reference, with the normalized beamforming gains at User 1, Eavesdropper 1, and Eavesdropper 2 being 0, $-36.1$ dB and $-34.5$ dB, respectively. The results demonstrate that the proposed precoder design algorithm effectively eliminates inter-user interference and achieves low gains at the eavesdropper positions, thereby enhancing the system's security.

\subsection{Received Constellations}
In this subsection, we explore the flexibility of the proposed algorithm by employing different modulation schemes for distinct users. The SINR for the legitimate users is fixed at $15$ dB, and other simulation parameters are consistent with those in Section V-A. The constellation diagrams for both the legitimate users and eavesdroppers are plotted by observing the total time slots of $K=800$. We examine a typical scenario where information bits towards the two distinct users adopt different modulations. Specifically, 8-PSK modulation is utilized for User 1, while 16 quadrature amplitude modulation (16-QAM) is adopted for User 2. Fig. \ref{Constellations} depicts the received constellations. Notably, the gain of the received symbols at the eavesdroppers are significantly lower compared to that of the legitimate users. Moreover, the constellations received by the eavesdroppers are completely scrambled, effectively preventing them from decoding the legitimate message, even in cases where they manage to amplify the received signals.

\begin{figure}[!t]
	\centering
	\subfloat[]{
		\label{berspaceus1}
		\includegraphics[width=0.25\textwidth]{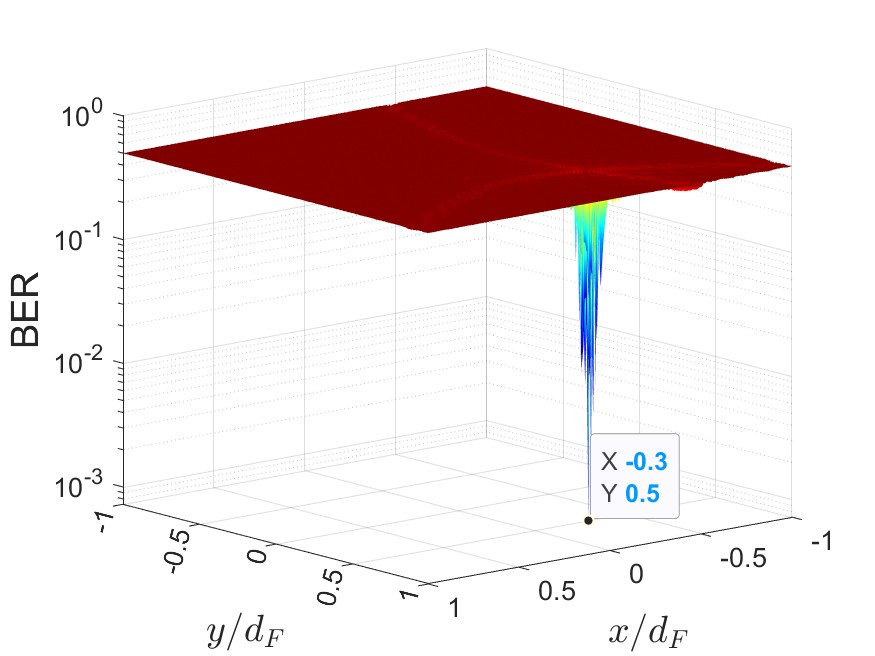}}
	\subfloat[]{
		\label{berspaceus2}
		\includegraphics[width=0.25\textwidth]{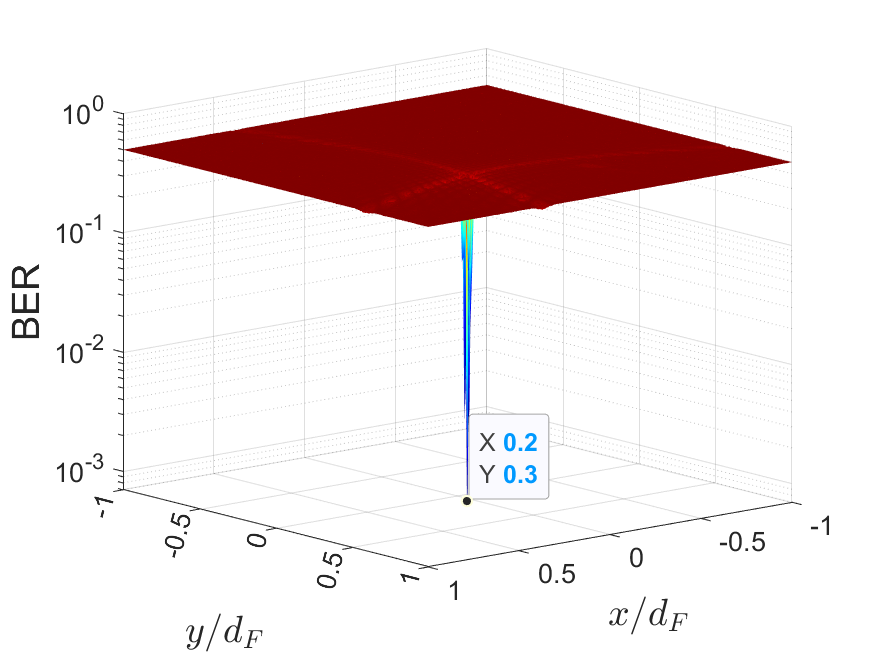}}
	\caption{BER performance at $z=0.55d_F$, where: (a) The BER of decoding information stream for User 1; (b) The BER of decoding information stream for User 2.}
	\label{berspace}
\end{figure}

\subsection{Performance Comparison in Terms of BER}
In this section, we assess the bit error rate (BER) for both legitimate users and eavesdroppers to evaluate the security efficacy of the proposed algorithm. The number of Monte-Carlo simulations is $100$, across a total of $1,000$ time slots. Utilizing quadrature phase shift keying (QPSK) modulation for all users, the theoretical BER at position ${\bf{r}}$ for the $m$th information stream is expressed as 
\begin{align}
	\text{BER}_\text{QPSK}({\bf{r}},m) = \mathcal{Q}\left( \sqrt {{\overline {{\rm{SINR}}} ({\bf{r}},m)}} \right) ,
\end{align}
where 
\begin{align}
\mathcal{Q}(x)=\frac{1}{{\sqrt {2\pi} }}\int_x^\infty {{e^{ - {\frac{1}{2}t^2}}}{\rm{d}}t} 
\end{align} 
represents the Q function.

\emph{1) BER Performance at $z=0.55d_F$:} Firstly, we explore the BER as a function of position in the plane at $z=0.55d_F$, and maintain the SINR for the legitimate users at a constant value of ${\beta _1^2}/{\sigma^2}={\beta _2^2}/{\sigma^2}=10$ dB. The transmission power is set to $P_t = 5{\left\| {{\bf{F}}{{\left( {{\bf{H}}_M^\dagger } \right)}^H}{{\bf{B}}_M}} \right\|_F^2}$. The simulation results are depicted in Fig. \ref{berspace}, with Fig. \ref{berspace}\subref{berspaceus1} illustrating the BER for decoding the information stream intended for User 1 and Fig. \ref{berspace}\subref{berspaceus2} for User 2. A detailed examination of the simulated BER surfaces reveals pronounced valleys at the positions of legitimate users. In contrast, the BER at any other position is higher than that at the locations of legitimate users, thereby ensuring the security performance in signal transmission to legitimate users.

\begin{figure}[!t]
	\centering
	\includegraphics[width=0.33\textwidth]{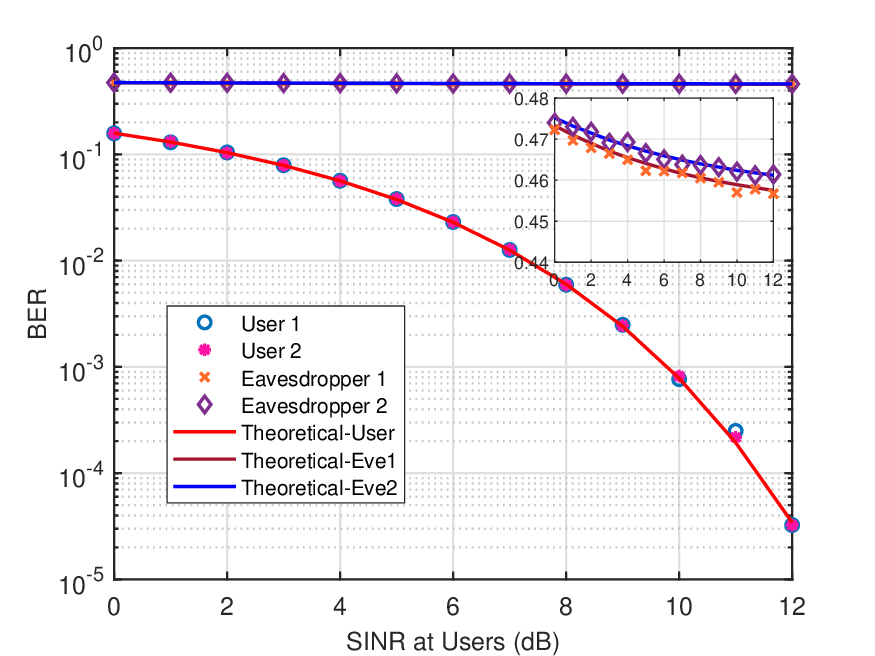}
	\caption{The BERs of legitimate users and eavesdroppers.}
	\label{ber1}
\end{figure}

\emph{2) Fixed Positions for Both Users and Eavesdroppers:} Moreover, we present the BER performance in a general case, with fixed users and eavesdroppers positions. We vary the SINR at the legitimate users, represented by ${\beta _m^2}/{\sigma^2}$ to evaluate the BER performance. We set $\beta _1=\beta _2$, and consider equal noise powers for the users and eavesdroppers. The comparative BER simulation results are presented in Fig. \ref{ber1}, which reveals that the BERs for legitimate users significantly decrease as the SINR increases. Conversely, the eavesdroppers experience substantially higher BERs compared to the legitimate users. Furthermore, the theoretical analysis perfectly characterizes the BERs of both the users and eavesdroppers.

\begin{table*}[]
	\scriptsize
     \renewcommand{\arraystretch}{1.5}
	\centering
	\caption {The positions of scatterers}
	\label{tab1}
	\begin{tabular}{c|c||c|c||c|c||c|c}
		\hline
		Scatter &Position&Scatter & Position &Scatter & Position &Scatter & Position\\ \hline
		1 & $(-0.3d_F,0.3d_F,0.8d_F)$ &2 & $(0.4d_F,0.5d_F,0.7d_F)$ &3 & $(0.4d_F,0.4d_F,0.55d_F)$ &4 & $(0.2d_F,0.2d_F,0.35d_F)$\\\hline
		5 & $(-0.1d_F,-0.6d_F,0.2d_F)$ &6 & $(0.55d_F,-0.6d_F,0.1d_F)$ &7 & $(-0.7d_F,-0.2d_F,0.1d_F)$ & -- & -----\\ \hline
	\end{tabular}
\end{table*}

\begin{figure}[!t]
	\centering
	\includegraphics[width=0.33\textwidth]{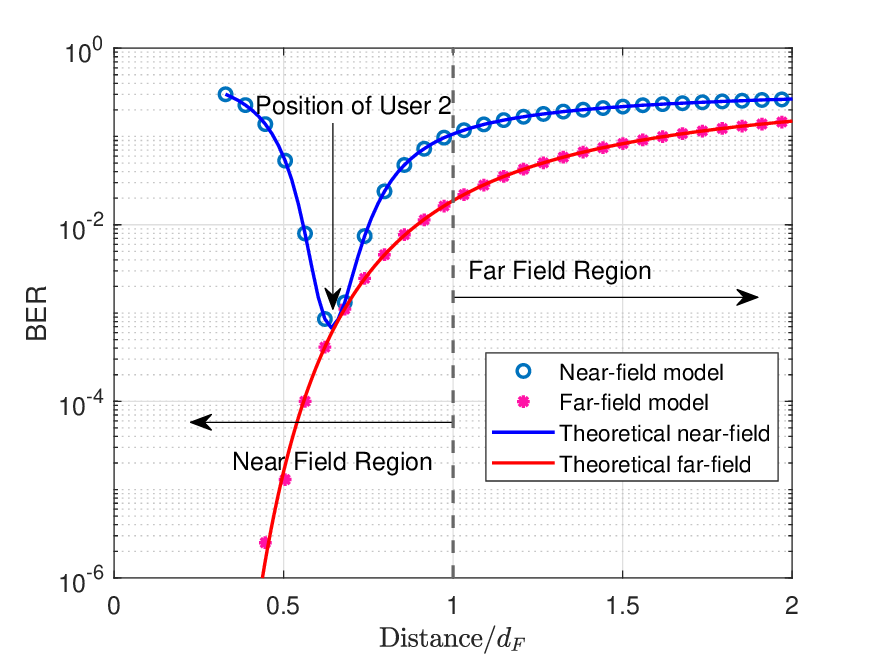}
	\caption{The BERs of Eavesdropper 2 with different channel models.}
	\label{bernearfar}
\end{figure}

\emph{3) Same Direction for User and Eavesdropper:} To further demonstrate the enhanced security provided by the proposed algorithm in near-field scenarios, we maintain the positions of User 1, User 2, and Eavesdropper 1, while replacing Eavesdropper 2 at $(0.1d_F,0.15d_F,0.275d_F)$. It is important to note that User 2 and Eavesdropper 2 are aligned in the same direction as BS. Then, we move Eavesdropper 2 along the same direction as User 2, which is illustrated in Fig. \ref{pos}\subref{samepos}. Under this condition, far-field direction modulation (DM) schemes fail to enable secure transmissions, as User 2 and Eavesdropper 2 share identical channel characteristics. We set the parameter $\xi$ to $2\times 10^{-6}$. The noise power is $-105$ dBm, and the SINR for legitimate users is fixed at $10$ dB. The BER simulation results of User 2 and Eavesdropper 2 are presented in Fig. \ref{bernearfar}. It is seen from Fig. \ref{bernearfar} that in the far-field channel model, Eavesdropper 2 has a lower BER than User 2 when it gets closer to the BS than the legitimate user, failing to guarantee secure transmission. This is because Eavesdropper 2 suffers from a lower path loss than User 2. In contrast, due to the chaotic constellation illustrated in Fig. \ref{Constellations}, Eavesdropper 2 maintains a high BER when it moves farther away from User 2 under the near-field channel model. When the eavesdropper is farther away from the BER than the legitimate user, the BER will degrade as expected, regardless of whether the near-field or far-field propagation model is used. However, the near-field model results in a higher BER for the eavesdropper compared to the far-field model. This is because the near-field model's unique spherical wave propagation concentrates energy on the legitimate user while reducing leakage to the eavesdropper. As the propagation distance from the BS increases, the performance gap between the two models narrows. These findings indicate that by leveraging the unique properties of the near-field channel, the proposed method is capable of achieving enhanced security at the precise locations of legitimate users.
\begin{figure*}[!t]
	\centering
	\subfloat[]{
		\label{nlospower}
		\includegraphics[width=0.33\textwidth]{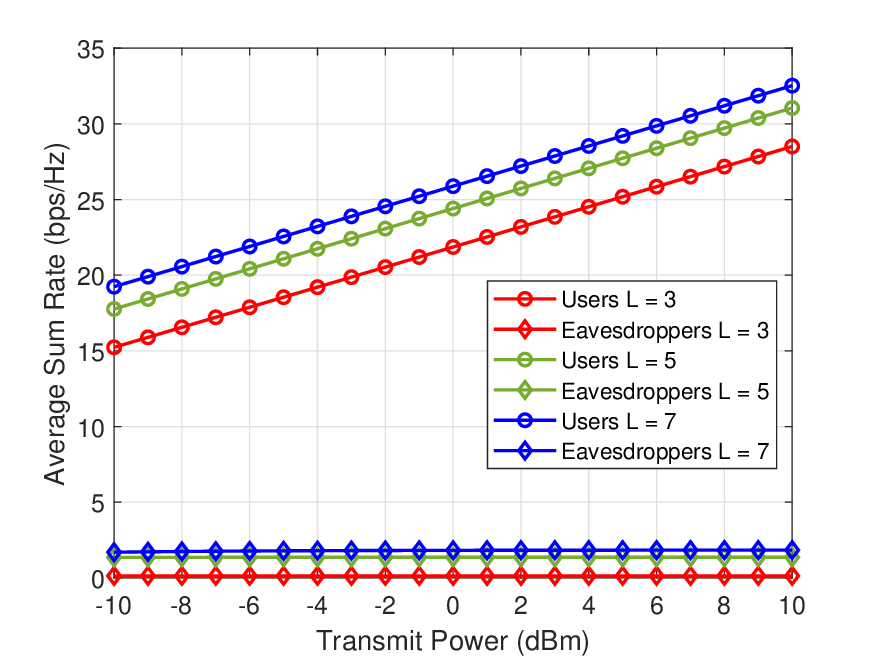}}
	\subfloat[]{
		\label{rate}
		\includegraphics[width=0.33\textwidth]{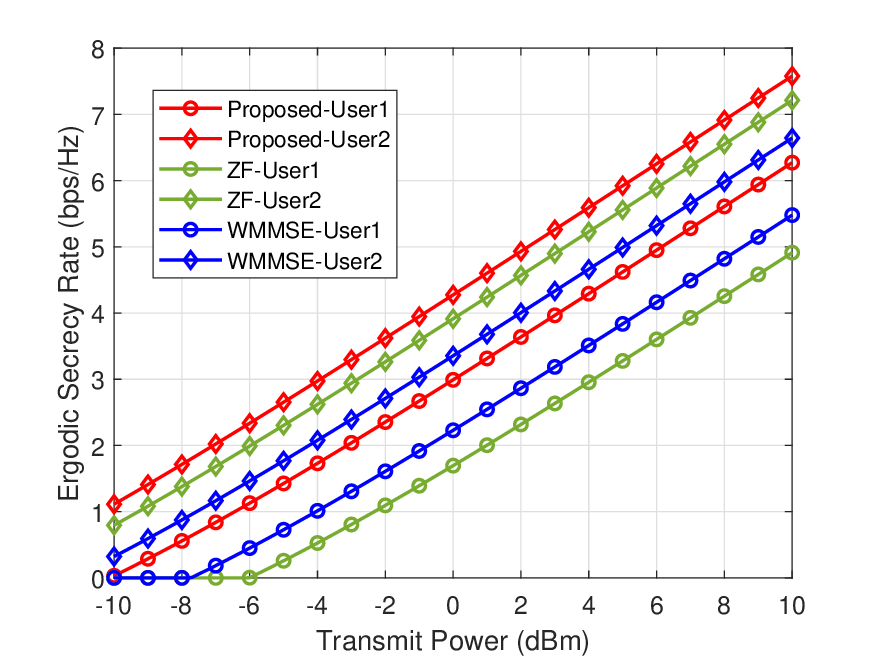}}
	\subfloat[]{
		\label{impcsi}
		\includegraphics[width=0.33\textwidth]{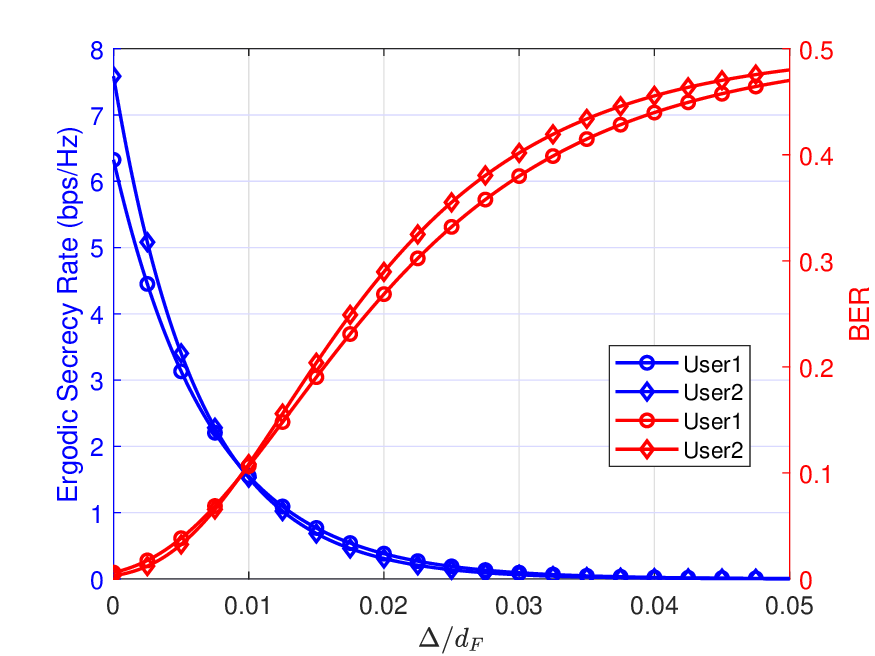}}\
	\subfloat[]{
		\label{hybrid}
		\includegraphics[width=0.33\textwidth]{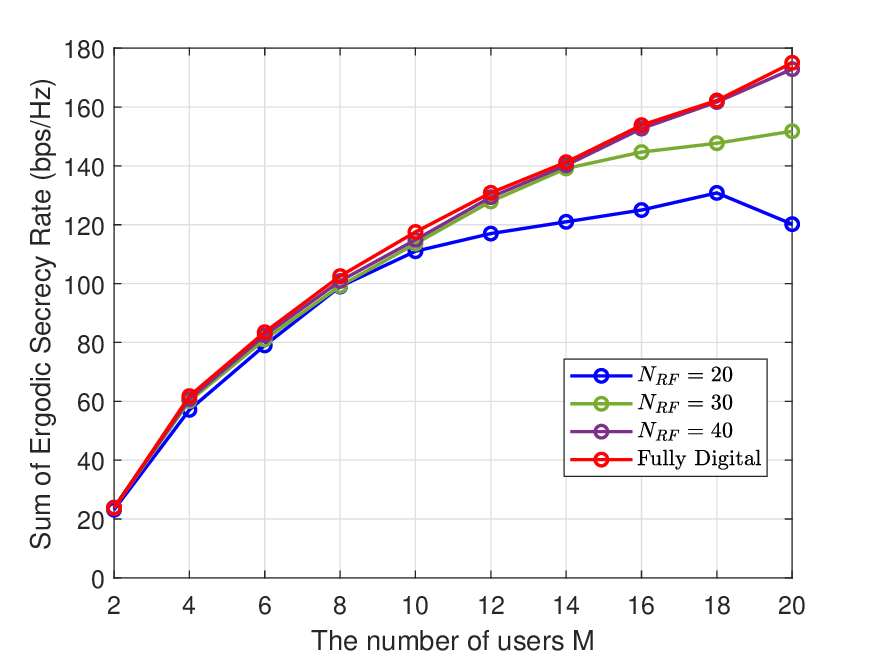}}
	\subfloat[]{
		\label{prob}
		\includegraphics[width=0.33\textwidth]{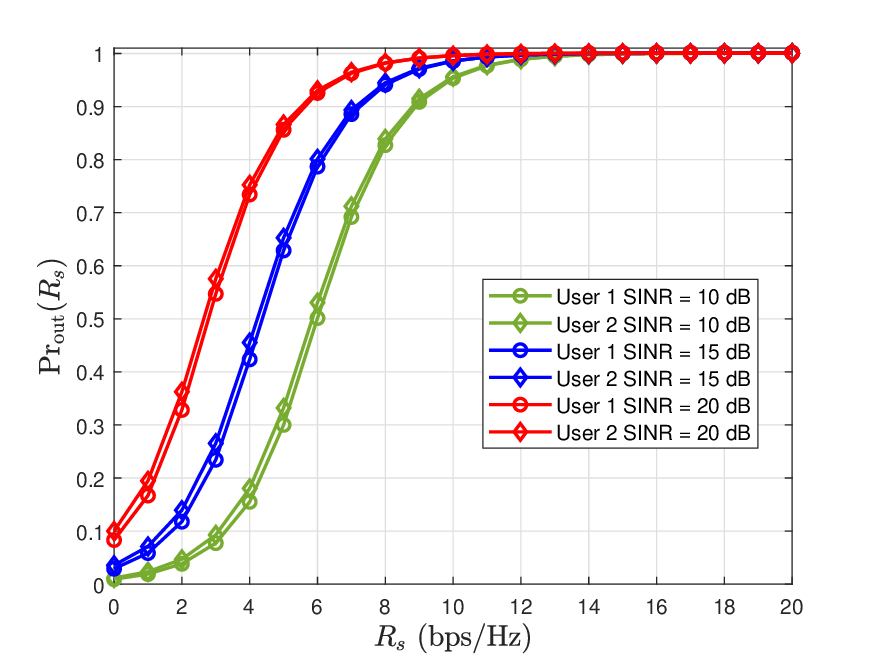}}
	\caption{(a) The average sum rate of legitimate users and eavesdroppers in multi-path channels; (b) The ergodic secrecy rate comparison of legitimate users; (c) The ergodic secrecy rate and BERs of legitimate users under imperfect CSI; (d) The sum of ergodic secrecy rate under random user and eavesdropper distribution; (e) The secrecy outage probability of legitimate users with different SINRs.}
	\label{figs}
\end{figure*}
\subsection{Sum Rate in Multi-path Channels}
Next, we examine the secure performance in multi-path propagation channels. The specific scatterer positions are listed in Table \ref{tab1}. We investigate the sum rate for legitimate users and eavesdroppers by changing the transmit power $P_t$ and the number of paths $L$. We set ${\bf{R}}=\sum\limits_{l = 1}^L {{\bf{h}}({{\bf{r}}^s_l})} {{\bf{h}}^H}({{\bf{r}}^s_l})$. Moreover, the noise power is set to $-105$ dBm, and $\xi = {0.9\sqrt P_t}/((1+\sqrt{ M })N_\text{RF}\sqrt {MN}) $. The sum rate performance is presented in Fig. \ref{figs}\subref{nlospower}. It is shown that under the same transmit power, the sum rates of the legitimate users increase with the number of paths. This is because, with perfect knowledge of CSI, multi-path diversity can be exploited by the BS to improve the rate. Furthermore, the sum rates for the legitimate users markedly increase as the transmit power increases. However, under identical transmit power conditions, the sum rates for the eavesdroppers are significantly lower than those of the legitimate users. These results effectively demonstrate that the proposed algorithm maintains secure wireless transmission even in multi-path channels.

\subsection{Secrecy Rate}
\emph{1) Secrecy Rate Compared with ZF and WMMSE Precoder:}
In this subsection, two baseline schemes, the ZF precoder and the WMMSE precoder are compared in terms of the ergodic secrecy rate for both legitimate users. The two eavesdroppers are moved to $(-0.15d_F,0.25d_F,0.275d_F)$ and $(0.1d_F,0.15d_F,0.275d_F)$, in the same direction of User 1 and User 2, respectively. The noise power and parameter $\xi $ are set to the same values as in Section V-D. We note that in Fig. \ref{figs}\subref{rate}, the secrecy rate for the legitimate users increases with the transmit power for all precoding schemes, with the proposed dynamic precoder exhibiting the highest secrecy rate at the same transmit power. Since User 2 is closer to the transmitting array, it has a higher secrecy rate than User 1.

{\emph{2) Secrecy Rate under imperfect CSI:}
To address the performance of the proposed method under imperfect CSI, we assume that the user location estimation error $\Delta{{\bf{r}}^u_m}$ satisfies
\begin{align}\label{errbound}
	\left\|\Delta{{\bf{r}}^u_m}\right\|\le\Delta.
\end{align}
Moreover, we set different $\Delta$ values and apply random $\Delta{{\bf{r}}^u_m}$ perturbations that satisfy \eqref{errbound} and follow a uniform distribution, while keeping the eavesdroppers positions consistent with those described in 1). 500 Monte-Carlo simulations are conducted to evaluate the performance of the proposed method. Fig. \ref{figs}\subref{impcsi} shows the ergodic secrecy rate and BER for User 1 and User 2 under different location estimation errors when $P_t=10$ dBm, $\sigma^2=-105$ dBm. As illustrated in Fig. \ref{figs}\subref{impcsi}, with the increase in the positioning error for the users, the ergodic secrecy rate for both users decreases, while the BER increases. This is because the beamfocusing becomes less accurate, leading to a lower SNR for both users. It should be noted that imperfect CSI primarily impacts the received signal quality at the intended users' locations. Due to the presence of AN, the eavesdropper still faces significant challenges in effectively decoding the information.}

\emph{3) Secrecy Rate under Random User and Eavesdropper Distribution:}
Next, we consider a more general scenario with different numbers of users randomly located in the near-field region, and each eavesdropper is randomly located within a circular area of radius $0.1d_F$ centered around the respective user. The transmit power and noise power are set to the same values as 2). Fig. \ref{figs}\subref{hybrid} shows the relationship between the number of users and the sum of ergodic secrecy rate when using hybrid precoding and fully digital precoding schemes. It can be seen that when the number of users is relatively small, the total ergodic secrecy rate increases monotonically with the number of users. Furthermore, when the number of RF chains exceeds twice the number of users, hybrid beamforming achieves performance comparable to fully digital beamforming. However, when the number of RF chains and users are both set to 20, the sum of the ergodic secrecy rate decreases. This drop occurs because there is insufficient null space to insert AN effectively.

\subsection{Secrecy Outage Probability}
In this subsection, we examine the secrecy outage probability of the two legitimate users. We set the SINR at the legitimate users to $10$ dB, $15$ dB, and $20$ dB, respectively. The corresponding secrecy outage probability curves relative to the threshold $R_s$ are illustrated in Fig. \ref{figs}\subref{prob}. It can be observed from Fig. \ref{figs}\subref{prob} that the secrecy outage probability increases with the threshold $R_s$, particularly when $R_s$ is less than $10$ bps/Hz. Furthermore, a higher SINR at the legitimate user results in a higher secrecy outage probability. This is because a higher SINR also improves the signal quality received by the eavesdroppers, increasing their opportunities of successfully decoding the information. 
\begin{figure*}[]
	\centering
	\subfloat[]{
		\label{2020}
		\includegraphics[width=0.33\textwidth]{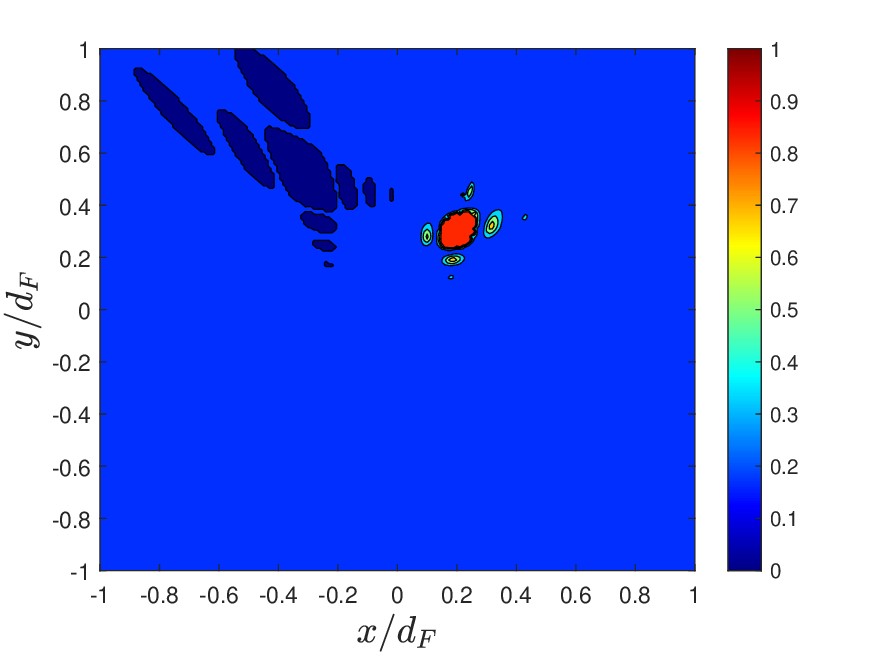}}
	\subfloat[]{
		\label{3030}
		\includegraphics[width=0.33\textwidth]{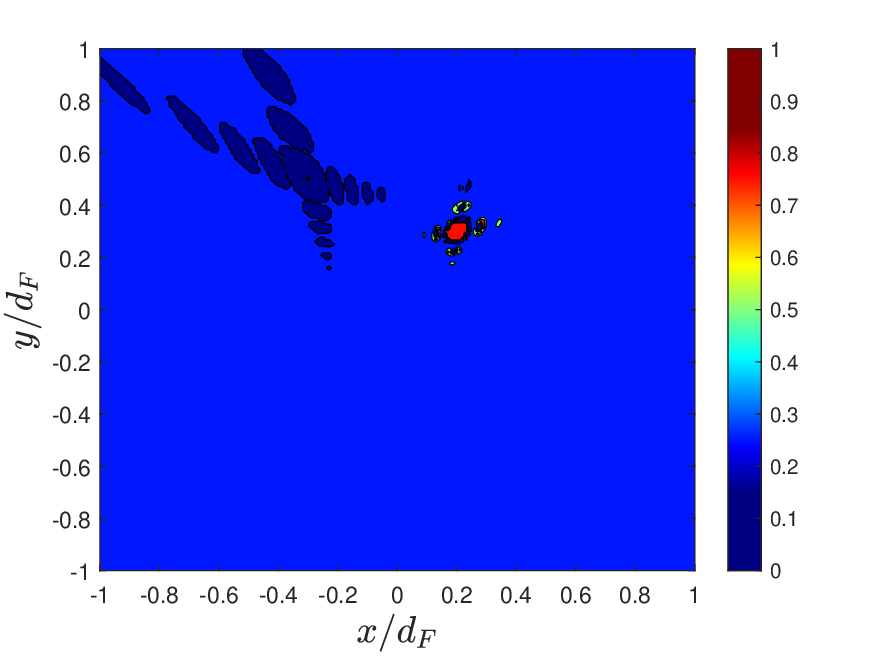}}
	\subfloat[]{
		\label{4040}
		\includegraphics[width=0.33\textwidth]{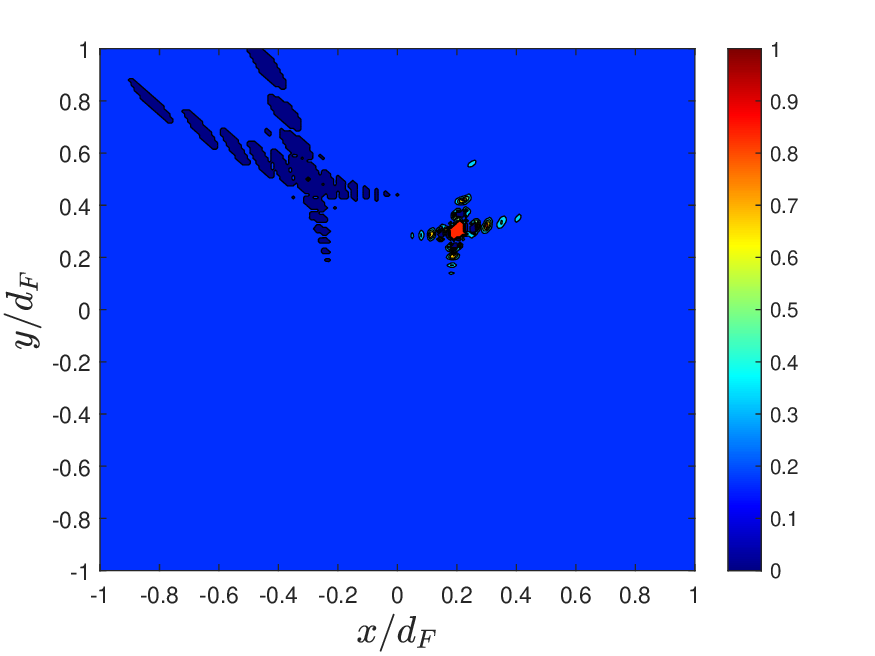}}
	\caption{Secrecy map at $z=0.55d_F$ of User 2 under different antenna number, where: (a) $20\times20$; (b) $30\times30$; (c) $40\times40$.}
	\label{map}
\end{figure*}

\subsection{Secrecy Map}
Finally, we calculate the secrecy outage probability at each point within the near-field region to generate a secrecy map. Using $20 \times 20$, $30 \times 30$, and $40 \times 40$ antenna arrays, we plot the secrecy outage probability for User 2 with respect to the eavesdropper at the plane $z = 0.55 d_F$, with $R_s = 5$ bps/Hz and SNR = 10 dB, under fully-digital transmission. As shown in Fig. \ref{map}, the secrecy outage probability near User 2 is close to 1, primarily due to the higher signal leakage in that region. In contrast, the secrecy outage probability in other areas is relatively low. Specifically, near User 1, the secrecy outage probability approaches 0, which is a result of ZF precoding eliminating inter-user interference, making it challenging for the eavesdropper to intercept User 2’s communication. As the number of antennas increases, the beamforming becomes more focused on User 2, thus reducing the region with a high secrecy outage probability. This emphasizes the crucial role of ELAA in enhancing the security of communication, particularly in the near-field regime.

\section{Conclusion}
In this paper, we introduced a dynamic hybrid beamforming architecture tailored for near-field multi-user PLS. Specifically, we designed the analog precoding based on the SVD of the wireless channel, and the symbol level baseband precoding was developed using AN-aided ZF, which not only adheres to power constraints but also effectively mitigates inter-user interference with a low computational complexity. This algorithm exploits the additional DoF provided by ELAA to distort the signal constellations perceived by passive eavesdroppers. Importantly, it supports a variety of modulation schemes, ensuring adaptivity to diverse legitimate user requirements and applicability in multi-path near-field channel environments. A thorough theoretical evaluation of the algorithm's effectiveness was conducted, examining metrics such as average SINR, achievable rate, secrecy capacity, secrecy outage probability, and the size of the secrecy zone. Numerical results confirm that the proposed method markedly enhances transmission security in both angular and distance dimensions, operating effectively without requiring eavesdroppers' CSI. In future work, we aim to extend the capabilities of our algorithm to near-field multi-user multiple-input multiple-output (MU-MIMO) scenarios to further enhance the security performance.

\begin{appendices}
	\begin{figure*}[t!]
		\begin{align}\label{wkg}
			{\bf{W}}(k){{\bf{g}}_m}{\bf{g}}_m^H{{\bf{W}}^H}(k) = {\bf{H}}_M^ \bot {{\bf{W}}_0}(k){{\bf{g}}_m}{\bf{g}}_m^H{\bf{W}}_0^H(k){({\bf{H}}_M^ \bot )^H} + {({\bf{H}}_M^\dagger )^H}{{\bf{B}}_M}{{\bf{g}}_m}{\bf{g}}_m^H{\bf{B}}_M^H{\bf{H}}_M^\dagger \nonumber\\
			+ {({\bf{H}}_M^\dagger )^H}{{\bf{B}}_M}{{\bf{g}}_m}{\bf{g}}_m^H{\bf{W}}_0^H(k){({\bf{H}}_M^ \bot )^H} + {\bf{H}}_M^ \bot {{\bf{W}}_0}(k){{\bf{g}}_m}{\bf{g}}_m^H{{\bf{B}}_M}{\bf{H}}_M^\dagger.\tag{72}
		\end{align}\hrulefill
	\end{figure*}
\section{Proof of Proposition 1}
By using $\left\| {\bf{X}} \right\|_F^2 = \text{tr}({{\bf{X}}^H}{\bf{X}})$, problem \eqref{problem1} can be rewritten as
\begin{subequations}
	\begin{align}
		&\max \quad \xi \\
		&\text{ s.t.}\quad A{\xi ^2} + B\xi + C \le 0\label{quardic}\\ 
		&\quad \quad \quad\xi >0\label{xipos}
	\end{align}
\end{subequations}
where $A$, $B$, and $C$ are given as \eqref{AA}-\eqref{CC}, respectively. Notice that $A>0$, the solution of quadratic inequality \eqref{quardic} is
\begin{align}
	\frac{{ - B - \sqrt {{B^2} - 4AC} }}{{2A}} \le \xi \le \frac{{ - B + \sqrt {{B^2} - 4AC} }}{{2A}}.
\end{align}
If and only if $C<0$, \eqref{quardic} has positive solutions to satisfy \eqref{xipos}. Consequently, the solution of problem \eqref{problem1} is
\begin{align}
	\xi = \frac{{ - B + \sqrt {{B^2} - 4AC} }}{{2A}},
\end{align}	
\setcounter{equation}{72}
which completes the proof.
\section{Proof of Proposition 2}
First, according to \eqref{wan}, ${\bf{W}}(k){\bf{g}}_m{\bf{g}}_m^H{{\bf{W}}^H}(k)$ can be calculated as \eqref{wkg}. Due to the fact that ${\varphi _{n_\text{RF},m}}(k) \sim \mathcal{U}\left( {0,2\pi } \right), \forall n \in\{1,2,...,N_\text{RF}\}, \forall m \in\{1,2,...,M\}$ are independent to each other at arbitrary pairs of time slots, we can obtain
\begin{align}
	\mathbb{E}\left[ {{e^{j{\varphi _{n_\text{RF},m}}(k)}}} \right] &= \frac{1}{{2\pi }}\int_0^{2\pi } {{e^{j{\varphi _{n_\text{RF},m}}(k)}}} d{\varphi _{n_\text{RF},m}}(k)=0 ,\label{meanej1}\\
	\mathbb{E}\left[ {{e^{-j{\varphi _{n_\text{RF},m}}(k)}}} \right] &= \frac{1}{{2\pi }}\int_0^{2\pi } {{e^{-j{\varphi _{n.m}}(k)}}} d{\varphi _{n_\text{RF},m}}(k)=0,\label{meanej2}
\end{align}
\begin{align}
	&\mathbb{E}\left[ {{e^{ - j{\varphi _{n_\text{RF},m}}(k)}}{e^{j{\varphi _{p,q}}(k)}}} \right]\nonumber\\ 
	=& \left\{ \begin{array}{l}
		1,\left( {n_\text{RF},m} \right) = \left( {p,q} \right)\\
		\mathbb{E}\left[ {{e^{ - j{\varphi _{n_\text{RF},m}}(k)}}} \right]\mathbb{E}\left[ {{e^{ j{\varphi _{p,q}}(k)}}} \right]= 0,\left( {n_\text{RF},m} \right) \ne \left( {p,q} \right)
	\end{array} \right..\label{meanej3}
\end{align}
Furthermore, recalling that the $(n_\text{RF},m)$-th entry of ${{\bf{W}}_0}(k)$ is ${w_{0,n_\text{RF},m}}(k) = \xi{e^{j{\varphi _{n_\text{RF},m}}(k)}},\forall n \in\{1,2,...,N_\text{RF}\}, \forall m \in\{1,2,...,M\}$, and according to \eqref{meanej1}, \eqref{meanej2}, we have
\begin{align}\label{w00}
	\mathbb{E}\left[ {{\bf{W}}_0}(k) \right] = {\bf{0}}_{N \times M},\quad \mathbb{E}\left[ {{\bf{W}}_0^H}(k) \right] = {\bf{0}}_{M \times N}.
\end{align}
Next, we calculate the mathematical expectation of the random terms in \eqref{wkg}. By adopting \eqref{meanej3}, for all ${\bf{g}}_m$, $m\in\{1,2,...,M\}$, 
\begin{align}\label{ex0}
	&\mathbb{E}\left[ {{\bf{W}}_0}(k){\bf{g}}_m{\bf{g}}_m^H{\bf{W}}_0^H(k) \right]
	=\mathbb{E}\left[{\bf{w}}_{0,m}(k){\bf{w}}_{0,m}^H(k) \right]
	={\xi^2}{\bf{I}}_N.
\end{align}
According to \eqref{w00}, \eqref{ex0}, and the fact that ${({\bf{H}}_M^ \bot )^H}={{\bf{H}}^\bot_M}$ and ${\bf{H}}_M^ \bot{\bf{H}}_M^ \bot= {\bf{H}}_M^ \bot$, we can obtain
\begin{align}
	\mathbb{E}\left[{\bf{H}}_M^ \bot {{\bf{W}}_0}(k){{\bf{g}}_m}{\bf{g}}_m^H{\bf{W}}_0^H(k){({\bf{H}}_M^ \bot )^H} \right]&={{\xi^2} {\bf{H}}_M^\bot },\label{ex1}\\
	\mathbb{E}\left[{({\bf{H}}_M^\dagger )^H}{{\bf{B}}_M}{{\bf{g}}_m}{\bf{g}}_m^H{\bf{W}}_0^H(k){({\bf{H}}_M^ \bot )^H}\right]&={\bf{0}}_{N \times N},\\	
	\mathbb{E}\left[{\bf{H}}_M^ \bot {{\bf{W}}_0}(k){{\bf{g}}_m}{\bf{g}}_m^H{{\bf{B}}_M}{\bf{H}}_M^\dagger\right]&={\bf{0}}_{N \times N},\label{ex2}
\end{align}

Finally, substituting \eqref{ex0} -- \eqref{ex2} into \eqref{wkg} yields
\begin{align}
	&	\mathbb{E}\left[{\bf{W}}(k){\bf{g}}_m{\bf{g}}_m^H{{\bf{W}}^H}(k)\right]\nonumber \\
	=&	{{\xi^2} {\bf{H}}_M^\bot } 
	+ ({{\bf{H}}^\dagger_M})^H{{\bf{B }}_M}{\bf{g}}_m{\bf{g}}_m^H{\bf{B }}_M^H{{\bf{H}}^\dagger_M},
\end{align}
which completes the proof.
\end{appendices}

\bibliographystyle{IEEEtran}
\bibliography{Reference}
\end{document}